\documentclass[prd,aps,preprint,superscriptaddress,showkeys,showpacs,
preprintnumbers,amsmath,amssymb,floatfix]{revtex4-1}
\usepackage{graphics,graphicx,color,amssymb,float}
\usepackage{epsfig}
\usepackage{amsfonts}
\usepackage{amsmath}
\usepackage{amssymb}
\usepackage[sort&compress]{natbib}
\usepackage{dcolumn}
\usepackage{multirow}
\usepackage{bigstrut}
\usepackage{mathrsfs}
\usepackage{type1cm}
\usepackage{placeins}
\usepackage{color}
\usepackage{rotating}
\usepackage{slashed}
\usepackage{subfigure}

\usepackage[mathscr]{euscript}
\newcommand{\bpi}{\boldsymbol{\pi}}
\newcommand{\bzeta}{\boldsymbol{\zeta}}

\newcommand{\Fp}{F_\pi}

\newcommand{\btau}{\boldsymbol{\tau}}

\newcommand{\bsig}{\boldsymbol{\sigma}}

\newcommand{\dslash}[1]{#1 \llap{/\kern-0.5pt}}

\newcommand{\boldpi}{\mbox{\boldmath $\pi$}}
\newcommand{\Dslash}[1]{#1 \llap{/\kern+1.2pt}}

\begin{document}

\preprint{\today}

\title{Tests of Lorentz and CPT symmetry with hadrons and nuclei}

\author{J. P. Noordmans}
\affiliation{Van Swinderen Institute for Particle Physics and Gravity, 
                  University of Groningen, 
                  9747 AG Groningen, The Netherlands}
\affiliation{CENTRA, Departamento de F\'isica, Universidade do Algarve, 8005-139 Faro, Portugal}

\author{J. de Vries}
\affiliation{Nikhef, Theory Group, Science Park 105, 1098 XG, Amsterdam, The Netherlands}
                   
\author{R. G. E. Timmermans}
\affiliation{Van Swinderen Institute for Particle Physics and Gravity, 
                  University of Groningen, 
                  9747 AG Groningen, The Netherlands}

\date{\today}
\vspace{3em}

\begin{abstract}
We explore the breaking of Lorentz and CPT invariance in strong interactions at low energy in the framework of chiral
perturbation theory. Starting from the set of Lorentz-violating operators of mass-dimension five with quark and gluon fields, we construct the effective chiral Lagrangian with hadronic and electromagnetic interactions induced by these operators. We develop the power-counting scheme and discuss loop diagrams and the one-pion-exchange nucleon-nucleon potential. The effective chiral Lagrangian is the basis for calculations of low-energy observables with hadronic degrees of freedom. As examples, we consider clock-comparison
experiments with nuclei and spin-precession experiments with nucleons in storage rings. We derive strict limits on the
dimension-five tensors that quantify Lorentz and CPT violation. 
\end{abstract}
\maketitle

\section{Introduction}
Lorentz symmetry~\cite{Ein05,Wig39,Wei89}, the covariance of the laws of physics under rotations and boosts in four-dimensional spacetime, plays a central role in physics and is at the basis of the standard model (SM) of particle physics and general relativity. In particle physics, it is closely related to the invariance under the combined transformations of charge conjugation, parity, and time reversal (CPT). In quantum field theories, with mild assumptions, Lorentz symmetry implies CPT invariance, while CPT violation implies Lorentz violation (LV)~\cite{Wes89,Gre02}. Nowadays, research into the breaking of Lorentz symmetry is strongly motivated by theories that attempt to unify quantum mechanics and general relativity~\cite{Lib13,Tas14}. Some of these theories contain mechanisms that naturally lead to Lorentz violation \cite{qgmodels}. The intriguing possibility exists that remnants of LV at high energy are detectable at energies that are in reach of present-day experiments. The detection of the corresponding signals would be a revolutionary discovery and could point us to the correct theory of quantum gravity. LV, in fact, is one of the few possibilities to get an experimental handle on quantum gravity.

In particle physics, the consequences of LV at low energy are conveniently studied within an effective field theory (EFT), which allows for a systematic and model-independent framework. The pertinent operators are built from SM fields coupled to fixed-valued Lorentz tensors (sometimes called ``background fields''), while keeping many desirable SM features, such as gauge invariance and the SM gauge-group structure, energy and momentum conservation, micro-causality, and observer Lorentz covariance \cite{sme}. The tensors parametrize LV, which presumably originates from more fundamental Lorentz-tensor fields that obtained a vacuum expectation value through spontaneous symmetry breaking at high energy.
This approach has led to the standard-model extension (SME) \cite{sme}, which is the most general and widely-used framework for theoretical and experimental considerations of Lorentz and CPT violation in particle physics.

At low energy, LV  results in unique experimental signals that are in principle easily distinguished from Lorentz-invariant physics beyond the SM, in particular frame dependence of observables and a dependence on sidereal time. Experimental constraints can be characterized and classified in terms of bounds on the components of the LV tensors in the SME. An overview of the existing experimental bounds can be found in Ref.~\cite{datatables}.
Most experimental bounds on LV have been obtained in the area of quantum electrodynamics, while recently progress also has been made in the weak sector~\cite{Noo13,Alt13,Dia13,Vos15}. However, most precision tests of Lorentz and CPT symmetry take place at low energies where quantum chromodynamics (QCD) is nonperturbative. This complicates the study of LV operators that contain quark or gluon fields, to the extent that only a relatively small number of direct bounds exists for the strong sector \cite{datatables}.

In this paper, therefore, we explore the use of chiral perturbation theory ($\chi$PT), the low-energy EFT of QCD~\cite{Wei79,Gas84} (for reviews, see e.g. Refs. \cite{Wei96,Ber06,Sch12}), to investigate the consequences of several higher-dimensional LV operators with quark and gluon fields. We construct, in Section~\ref{sec:LVlagr}, chiral Lagrangians that describe LV interactions between pions, nucleons, and photons. The large nucleon mass is treated in the heavy-baryon approach \cite{Jen91}. Our approach is similar in spirit to previous studies of the breaking of parity~\cite{Kap93} and time reversal~\cite{Mer10} from dimension-six operators, as applied, for example, to P-odd~\cite{Mae00} and P- and T-odd~\cite{Hoc05} electromagnetic form factors of the nucleon. Within this framework, it becomes possible to study various LV observables for hadronic and nuclear degrees of freedom. In Section \ref{sec:LVham} we first construct the LV Hamiltonian, and next we identify in Section \ref{sec:Expts} observables for clock-comparison experiments with nuclei in atoms and ions and storage-ring experiments with nucleons. We obtain bounds on our LV tensors from existing experiments and identify opportunities to further constrain the parameter space. We end with a summary and outlook in Section~\ref{sec:Summary}. In Appendix~\ref{app:so4} we briefly review the construction of the chiral Lagrangian and the use of naive dimensional analysis. Appendix~\ref{app:eom} is devoted to the use of field redefinitions to reduce the number of effective operators.

\section{The Lorentz-violating chiral Lagrangian}
\label{sec:LVlagr}
\subsection{Operators with quarks and gluons}
We start with a set of operators relevant below the electroweak scale $\Lambda_{\rm F}\simeq 250$ GeV, but above the scale of chiral-symmetry breaking $\Lambda_\chi\simeq 2\pi F_\pi\sim$ 1 GeV in QCD, where $F_\pi\simeq 185$ MeV is the pion decay constant. LV is associated with a high-energy scale $\Lambda_{\rm LV}$ beyond $\Lambda_{\rm F}$, presumably to be identified with the Planck scale. Many LV operators have been discussed elsewhere in the literature. In Ref.~\cite{sme} all possible LV operators compatible with the SM gauge structure and of mass-dimension 3 and 4 are given. This restriction to power-counting renormalizable operators is sometimes called the minimal Standard-Model Extension (mSME). A characterization of nonminimal, higher- (5-, 6-,$\,\ldots$) dimensional operators exists for electrodynamics, neutrinos, and free fermions \cite{Kos09}.

In an EFT framework, higher-dimensional operators are expected to be suppressed by powers of some high-energy scale, $\Lambda_{\rm UV}$. In this respect, dimension-3 and -4 operators are less natural in an EFT for LV, where one assumes that $\Lambda_{\rm UV}=\Lambda_{\rm LV}$. Additional symmetry arguments are then needed to prevent the LV physics at high energy from resulting in large dimension-3 and -4 LV operators at a low-energy scale such as $\Lambda_\chi$. To evade the strong experimental limits on LV, these symmetry arguments should forbid the appearance of the dimension-3 and -4 operators, or at least make them scale like $\Lambda_{\rm IR}^2/\Lambda_{\rm UV}$ and $\Lambda_{\rm IR}/\Lambda_{\rm UV}$ respectively, where $\Lambda_{\rm IR}$ is e.g. the scale of supersymmetry breaking, $\Lambda_{\rm SUSY}$. Remarkably, for LV in the minimal supersymmetric standard model, the lowest dimension for LV operators is 5~\cite{Gro05,Bol05,Mat08}, so that LV is suppressed by at least one power of the high-energy scale. A similar suppression of dimension-3 and -4 operators occurs when we construct the effective chiral Lagrangian that is induced by dimension-5 operators in the LV QCD Lagrangian. 

In Ref.~\cite{dim5lagrangian} all dimension-5 operators were classified that can be built out of SM fields and are restricted by a set of ``UV-safety'' conditions that protect the operators from transmuting into lower-dimensional operators by quantum effects. In this paper, we will restrict ourselves to a subset of the quark and gluon operators listed in Ref.~\cite{dim5lagrangian}. The operators we choose are, for our exploratory purpose, the most interesting ones from the point of view of $\chi$PT. Other LV QCD operators can be treated in the same way, but we leave this for future work. 

At a scale of 1 GeV there is a limited set of protected dimension-5 operators in the quark sector. They are summarized in Eq.~(18) of Ref.~\cite{dim5lagrangian}. Of this set, we consider the only two that explicitly contain the gluon field strength $G^{\mu\nu} = t^a G^{a,\mu\nu}$ ($t^a=\frac{1}{2}\lambda^a$, $a=1,\dots,8$, where $\lambda^a$ are the Gell-Mann matrices, are the generators of the $SU(3)$ color group).  They are given by the Lagrangian density
\begin{equation}
\mathcal{L}_q^{\rm LV} = \sum_{q=u,d}\left[ C^q_{\mu\nu\rho}\bar{q}\gamma^\mu G^{\rho\nu} q + D^q_{\mu\nu\rho}\bar{q}\gamma^\mu\gamma^5 G^{\rho\nu}q\right]\ .
\label{lagr1gevQ}
\end{equation}
Both operators also violate CPT invariance. Our naming of the LV tensors differs from Ref.~\cite{dim5lagrangian}, wherin $C^q$ is called $D_{qg}$, while $D^q$ is $D_{qg,5}$. Although these operators should be considered as part of a theory where the $W$ and $Z$ bosons are already integrated out, one should keep in mind that $C$ and $D$ contain (different) contributions from the same high-energy operators, because of mixing due to $W$- and $Z$-boson loops. Lacking a renormalization-group analysis for such operators, we will here consider $C$ and $D$ to be independent.
For isospin considerations we split the Lagrangian density in Eq.~\eqref{lagr1gevQ} in two parts,
\begin{eqnarray}
\mathcal{L}_q^{\rm LV} &=& C^+_{\mu\nu\rho}\bar{Q}\gamma^\mu G^{\rho\nu}Q + D^+_{\mu\nu\rho}\bar{Q}\gamma^\mu \gamma^5 G^{\rho\nu}Q \nonumber \\
 && + C^-_{\mu\nu\rho}\bar{Q}\gamma^\mu G^{\rho\nu}\tau_3 Q + D^-_{\mu\nu\rho}\bar{Q}\gamma^\mu \gamma^5 G^{\rho\nu}\tau_3 Q \ ,
\label{lagr1gevsplit}
\end{eqnarray}
with $Q= (u \, d)^T$, $X^{\pm}_{\mu\nu\rho} = (X^u_{\mu\nu\rho} \pm X^d_{\mu\nu\rho})/2$ for $X\in\{C,D\}$,  and $\tau_3$ the third Pauli matrix.

Operators similar to those in Eq.~\eqref{lagr1gevQ} exist that contain the photon field strength $F^{\mu\nu}$ instead of the gluon field strength. Some phenomenological effects of these operators are considered in Refs.~\cite{Bol08,Sta14}. Since we are interested in observables for non-strange baryons, we have focused on operators with up and down quarks only. Our analysis can be extended to include the strange quark and observables with kaons and hyperons. 

In addition to the quark operators we consider the only dimension-5 pure-gauge term that satisfies the UV-safety conditions of Ref.~\cite{dim5lagrangian}. It is given by the Lagrangian density
\begin{equation}
\mathcal{L}_g^{\rm LV} = H_{\mu\nu\rho}\,{\rm Tr}\left(G^{\mu\lambda}D^\nu\tilde{G}^{\rho}_{\;\;\;\lambda}\right)\ ,
\label{lagr1gevG}
\end{equation}
where $\tilde{G}^{\mu\nu}=\frac{1}{2}\varepsilon^{\mu\nu\rho\sigma}G_{\rho\sigma}$ is the dual tensor of $G^{\mu\nu}$. In Ref.~\cite{dim5lagrangian} $H^{\mu\nu\rho}$ is called $C_{{\rm SU}(3)_C}^{\mu\nu\rho}$.

\begin{table}[t!]
\centering
\begin{tabular}{cccc}
\hline\hline
& $C_{\mu\nu\rho}$ & $D_{\mu\nu\rho}$ & $H_{\mu\nu\rho}$ \\
\hline
P & $(-1)^{\mu}(-1)^{\nu}(-1)^{\rho}$ & $-(-1)^{\mu}(-1)^{\nu}(-1)^{\rho}$ & $-(-1)^{\mu}(-1)^{\nu}(-1)^{\rho}$ \\
T & $-(-1)^{\mu}(-1)^{\nu}(-1)^{\rho}$ & $-(-1)^{\mu}(-1)^{\nu}(-1)^{\rho}$ & $(-1)^{\mu}(-1)^{\nu}(-1)^{\rho}$ \\
C & $+1$ & $-1$ & $+1$ \\
\hline\hline
\end{tabular}
\caption{The transformation properties of the LV operators contracted by the tensors $C$, $D$, and $H$ under $C$, $P$, and $T$. $(-1)^\mu$ is equal to $+1$ ($-1$) if $\mu$ is a time-like (space-like) index.}
\label{tab:cptproperties}%
\end{table}

The real tensor components $C^q_{\mu\nu\rho}$ and $D^q_{\mu\nu\rho}$ describe LV in the quark-gluon interactions, whereas $H^{\mu\nu\rho}$ parametrizes the LV of gluonic interactions. All the LV tensor components have mass-dimension $-1$. Constraints on the symmetry of the components are derived from UV-safety considerations in Ref.~\cite{dim5lagrangian}: $X_{\mu\nu\rho} = \tfrac{1}{2}(X_{\mu\nu\rho}+ X_{\rho\nu\mu})$,  with $X\in\{C,D\}$, while $H_{\mu\nu\rho}$ is fully symmetric in all its Lorentz indices. Additionally, all traces of the LV tensors vanish. Due to these symmetries there are 16 independent components of $H_{\mu\nu\rho}$, while the observable parts of $C$ and $D$, i.e. $X_{\mu[\nu\rho]} = \tfrac{1}{2}(X_{\mu\nu\rho}- X_{\mu\rho\nu})$, each also have 16 independent components. The transformation properties of the LV operators under the discrete-symmetry transformations $C$, $P$, and $T$ are summarized in Table~\ref{tab:cptproperties}.

\subsection{Operators with nucleons and pions}
\label{Npioperators}
At momenta $p$ of order of the pion mass $p \sim m_\pi \ll \Lambda_\chi \sim 1$ GeV, the above operators induce interactions among the relevant low-energy degrees of freedom, pions ($\pi$), nucleons ($N$), and photons ($A_\mu$). To derive these interactions we employ $\chi$PT~\cite{Wei79,Gas84}. The standard $\chi$PT Lagrangian contains all interactions allowed by the QCD symmetries. In the limit of zero up- and down-quark masses and charges, the QCD Lagrangian has an $SU(2)_L\times SU(2)_R \sim SO(4)$ chiral symmetry. Chiral symmetry is spontaneously broken to its $SO(3)$ isospin subgroup, resulting in a triplet of (almost) massless Goldstone bosons, the pions. In this limit, pions only interact via space-time derivatives, allowing for the calculation of hadronic observables in perturbation theory, with expansion parameter $p/\Lambda_\chi$, where $p$ is the typical momentum of the process under consideration. The pion fields can be parametrized in infinitely many ways. We use stereographic coordinates \cite{Wei96}, as reviewed briefly in Appendix~\ref{app:so4}, but different  choices give identical results. For a generalization to $SU(3)_L\times SU(3)_R$ the standard formalism reviewed in Ref.~\cite{Sch12} would be indicated.

Although the operator form of the effective hadronic interactions is dictated by symmetry considerations, each interaction is multiplied by a low-energy constant (LEC) that parametrizes the nonperturbative dynamics. The values of these LECs do not follow from symmetry arguments alone. In principle these LECs can be calculated with lattice QCD, but for the LV cases discussed here this has not been done. Alternatively, if a Lorentz- or CPT-violating signal would be detected, the LV LECs can be fitted to the experimental data. In the absence of such LV signals, we resort to naive dimensional analysis (NDA)~\cite{Man84}, cf. Appendix~\ref{app:so4}, to estimate the LECs at the order-of-magnitude level.

The chiral (and gauge) symmetries are incorporated with covariant derivatives for the pion,
\begin{equation}
(D_\mu \boldsymbol{\pi})_a = D^{-1}(\partial_\mu \delta_{ab} + eA_\mu \epsilon_{3ab}) \pi_b\ ,
\end{equation}
and for the nucleon,
\begin{equation}
\mathcal{D}_\mu N = \left(\partial_\mu + \frac{i}{F_\pi^2}\boldsymbol{\tau}\cdot\boldsymbol{\pi}\times D_\mu \boldsymbol{\pi} + \frac{ie}{2} A_\mu(1+\tau_3)\right)N\ ,
\end{equation}
where $D= 1+\boldsymbol \pi^2/F_\pi^2$, $e>0$ is the proton charge, $\boldsymbol{\tau}$ are the Pauli isospin matrices, and $a,b$ are isospin indices. The low-energy effective Lagrangian involves an infinite number of interactions ordered by the expected size of their contributions to physical processes. Each effective interaction is associated with a chiral index \cite{Wei79,Gas84}
\begin{equation}
\Delta = d+f/2-2 \ ,
\label{Delta}
\end{equation}
where $d$ counts the number of (covariant) derivatives and $f$ the number of nucleon fields appearing in the interaction. Because $m_N/\Lambda_\chi$ is not a small number, time derivatives acting on nucleon fields are not suppressed. However, the combination $(i\Dslash {\mathcal D} -m_N)$ is still small and increases $d$ by one. The leading terms in the chiral-symmetric Lagrangian (that is, with the lowest chiral index $\Delta =0$) are then given by
\begin{eqnarray}\label{LO1}
{\mathcal L}_\chi^{\Delta=0} &=& 
\frac{1}{2} D_{\mu} \boldpi \cdot D^{\mu} \boldpi + \bar{N}\left( \Dslash {i\mathcal D}-m_N
         -\frac{ g_A}{F_\pi}( \btau\cdot D_{\mu} \boldpi)\gamma^{\mu}\gamma_5\right) N\ ,
\end{eqnarray}
in terms of the nucleon mass $m_N$ and the axial-vector coupling $g_A\simeq 1.27$. 

Chiral symmetry is broken by the masses of the up and down quarks, but, being small, these can be incorporated in the expansion by letting $d$ increase by two for each quark-mass insertion. The most important consequence is that the pion acquires a small mass through
\begin{equation}\label{LO2}
{\mathcal L}^{\Delta=0}_{m_\pi} = -\frac{m_\pi^2}{2D}\boldpi^2\,\,\,.
\end{equation}

In a similar fashion we can construct the hadronic interactions induced by the LV operators in Eqs.~\eqref{lagr1gevsplit} and \eqref{lagr1gevG}. We assume that there arises no additional Lorentz or CPT violation from the QCD phase transition itself, such that the symmetry properties of the LV coefficients remain intact when going from the quark-gluon to the $\chi$PT Lagrangian. Although all operators in Eqs.~\eqref{lagr1gevsplit} and \eqref{lagr1gevG} break Lorentz symmetry, they transform differently under chiral symmetry. The operator in Eq.~\eqref{lagr1gevG} and the first two terms in Eq.~\eqref{lagr1gevsplit} are invariant under global $SU_L(2)\times SU_R(2)$ chiral transformations, and therefore  induce low-energy interactions that are chiral invariant as well. The interactions, however, have different symmetrization properties of the Lorentz indices as well as different properties under the individual discrete-symmetry transformations $C$, $P$, and $T$ (see Table~\ref{tab:cptproperties}). Therefore, they lead to different chiral-invariant interactions at lower energies. 

In contrast, the last two terms in Eq.~\eqref{lagr1gevsplit} break chiral symmetry explicitly and thus induce chiral-breaking hadronic interactions. In particular, they give rise to operators that involve pion fields without the spacetime derivatives that are necessary for chiral-invariant interactions~\cite{Wei96}. The chiral operators resulting from the $C^-_{\mu\nu\rho}$ and $D^-_{\mu\nu\rho}$ terms can be easily constructed by noticing that the corresponding operators transform as, respectively, the $34$ and $12$ components of the antisymmetric $SO(4)$ tensor
\begin{equation}\label{eq:tensors}
T^{\mu\rho\nu}=\left(\begin{array}{cc}
\varepsilon_{abc}\bar Q \gamma^{\mu}\gamma^5 \tau_c \,G^{\rho\nu} Q & \bar Q \gamma^{\mu} \tau_a \,G^{\rho\nu}Q\\
-\bar Q \gamma^{\mu} \tau_a \,G^{\rho\nu} Q             &0
\end{array} \right).
\end{equation}

As we discuss below, the strongest experimental constraints result from LV two-point interactions for the nucleon. These two-point interactions are induced by the LV tensors $C^{\pm}_{\mu\nu\rho}$ and $H^{\pm}_{\mu\nu\rho}$. At the level of pions and nucleons the former give rise to the operators 
\begin{subequations}
\begin{eqnarray}
\mathcal{L}_{\chi C^+} &=& \frac{i}{m_N}\tilde{C}^{+}_{\mu\nu\rho}\bar{N}\sigma^{\nu\rho}\mathcal{D}^\mu N + \mathrm{H.c.} \label{chirallagrCplus} \\
\mathcal{L}_{\chi C^-} &=& \frac{i}{m_N}\tilde{C}^{-}_{\mu\nu\rho}\bar{N}\left[\tau_3 - \frac{2}{F_\pi^2 D}\left(\bpi^2\tau_3  - \pi_3 \btau\cdot\bpi\right)\right]\sigma^{\nu\rho}\mathcal{D}^\mu N + \mathrm{H.c.} \ , \label{chirallagrCmin}
\end{eqnarray}
\label{chirallagrC}%
\end{subequations}
where H.c. means hermitian conjugate. We denote the LV LECs at the hadronic level with a tilde. The LV components $\tilde{C}^{\pm}_{\mu\nu\rho}$ are related to $C^\pm_{\mu\nu\rho}$ by $\tilde{C}^{\pm}_{\mu\nu\rho} = c^\pm C^{\pm}_{\mu\nu\rho}$, where $c^\pm = \mathcal O ( \Lambda_\chi \Fp )$ is a strong-interaction matrix element estimated with NDA \cite{Man84}. We introduce a factor of $1/m_N$ for each covariant nucleon derivative to keep the time derivatives from spuriously lowering the chiral index of the operators, given by $\Delta = -1$ for the dominant terms in Eqs.~\eqref{chirallagrC}.

Chiral symmetry relates the nucleon-nucleon ($N\!N$) operators to pion-nucleon ($\pi N$) interactions. However, the strongest constraints result from the terms without pions. Operators of different form exist at this order, but in Appendix~\ref{app:eom} we show that these are redundant. In all hadronic interactions we also omit terms with additional nucleon covariant derivatives, because by using the equations of motion such terms can be reduced to operators of the same form plus higher-order terms. The form of the free-nucleon operators in Eqs.~\eqref{chirallagrC} agrees with the effective operator for $D_{qg}$ obtained in Ref.~\cite{dim5lagrangian}.

Similar to $C^+_{\mu\nu\rho}$, the $H_{\mu\nu\rho}$ operator induces only contributions to the nucleon two-point function at this order, viz.
\begin{eqnarray}
\mathcal{L}_{\chi H} &=&  \frac{1}{m_N^2}\tilde{H}_{\mu\nu\rho}\bar{N}\gamma^\mu \gamma^5 \mathcal{D}^\nu \mathcal{D}^\rho N +\mathrm{H.c.}\ ,
\label{chirallagrH}
\end{eqnarray}
with the LV LEC $\tilde{H}_{\mu\nu\rho}= h H_{\mu\nu\rho} $,  where $h = \mathcal O(\Lambda_\chi^2)$ is a strong-interaction matrix element estimated by NDA. Redundant terms are again discussed in Appendix~\ref{app:eom}.

In contrast to the $C$ and $H$ tensors, the tensors $D^{\pm}_{\mu\nu\rho}$ do not lead to a nucleon two-point function at any chiral order. In fact, at lowest order ($\Delta = -1$) only $D^-$ contributes. The relevant Lagrangian is given by 
\begin{eqnarray}
\mathcal{L}_{\chi D^-} &=& \frac{i}{m_N F_\pi D} \tilde{D}^-_{\mu\nu\rho}\bar{N}(\boldsymbol{\tau}\times\boldsymbol{\pi})_3\sigma^{\nu\rho}\mathcal{D}^\mu N   + \mathrm{H.c.} \label{chirallagrDmin}
\end{eqnarray}
The LV LEC is again defined by $\tilde{D}^{-}_{\mu\nu\rho} = d^-D^-_{\mu\nu\rho}$, with the strong-interaction matrix element $d^- = \mathcal O ( \Lambda_\chi \Fp )$ according to NDA. Because $C^-_{\mu\nu\rho}$ and $D^-_{\mu\nu\rho}$ are components of the same $SO(4)$ tensor, chiral symmetry gives the relation $d^- = 2c^-$. Additional redundant operators are discussed in Appendix~\ref{app:eom}. The leading terms for $D^+$, with chiral index $\Delta = 0$, read
\begin{eqnarray}\label{chirallagrDplus}
\mathcal{L}_{\chi D^+} &=& \frac{1}{m_N^2 F_\pi}\breve{D}^+_{\mu\nu\rho\alpha\beta}\bar{N}(\boldsymbol{\tau}\cdot D^\mu \boldsymbol{\pi})\sigma^{\nu\rho}\mathcal{D}^\alpha \mathcal{D}^\beta N + {\rm H.c.}\ ,
\end{eqnarray}
with $\breve{D}^+$ given by
\begin{equation}\label{Dbrevedef}
\breve{D}^+_{\mu\nu\rho\alpha\beta} = \tilde{D}^{+,1}_{\mu\rho\nu}g_{\alpha\beta} + \tilde{D}^{+,2}_{\alpha\nu\rho}g_{\mu\beta} + \tilde{D}^{+,3}_{\alpha[\beta\rho]}g_{\mu\nu} \ ,
\end{equation}
and the LV LECs defined as $\tilde{D}^{+,i}_{\mu\nu\rho} = d^+_i D^+_{\mu\nu\rho}$ with $d^+_i = \mathcal{O}(\Fp)$. The metric tensor $g_{\alpha\beta}$ in the first term of $\breve{D}^+$ contracts two covariant derivatives. Since at lowest order $\mathcal{D}^2 N = -m_N^2 N$, we see that it represents the simple operator $\tilde{D}^{+,1}_{\mu\nu\rho}\bar{N}(\boldsymbol{\tau}\cdot D^\mu \boldsymbol{\pi})\sigma^{\nu\rho}N$.

The operators in Eqs.~\eqref{chirallagrDmin} and ~\eqref{chirallagrDplus} will induce loop corrections to the nucleon Lagrangian. We will see an example of this in Section~\ref{sec:pionloop}. However, since nucleon two-point functions are not allowed by the symmetries of the original operators, they will also not be induced by quantum effects at first order in LV. It turns out that the dominant observable effects of the loop corrections are represented by nucleon two-point functions coupled to the electromagnetic field strength. At leading order, such operators have to take the form (see Appendix~\ref{app:eom})
\begin{eqnarray}
\mathcal{L}_{\chi DF}  &=& \frac{e}{m_N^3}\bar{N}\breve{D}^F_{\alpha\beta\mu\nu\rho\sigma\lambda}\gamma^5\sigma^{\alpha\beta} \mathcal{D}^\rho \mathcal{D}^\sigma \mathcal{D}^\lambda N F^{\mu\nu}+\mathrm{H.c.}+ \dots \ ,
\label{effDlagr}
\end{eqnarray}
where $\breve{D}^F$ is an isospin matrix analogous to Eq.~\eqref{Dbrevedef} and the dots represent $\pi N$ interactions that chiral symmetry relates to the displayed operator. The tensor $\breve{D}^F$ is built from the LV components $D^{\pm}_{\mu\nu\rho}$, $\tau_3$, the metric tensor, and low-energy constants of order one. This results in many inequivalent contributions to $\breve{D}^F$, each of which has its own LEC. It goes beyond the scope of this work to list them all, but two relevant examples are
\begin{subequations}
\begin{eqnarray}
\breve{D}^F_{\alpha\beta\mu\nu\rho\sigma\lambda} &\ni& g_{\beta\rho}g_{\sigma\lambda}(\tilde{D}^{F^+,1}_{\alpha\mu\nu}+\tau_3 \tilde{D}^{F^-,1}_{\alpha\mu\nu})\ , \label{simplestF} \\
\breve{D}^F_{\alpha\beta\mu\nu\rho\sigma\lambda} &\ni& g_{\alpha\rho}g_{\nu\sigma}(\tilde{D}^{F^+,2}_{\lambda[\beta\mu]}+ \tau_3 \tilde{D}^{F^-,2}_{\lambda[\beta\mu]})\ . \label{loopcorrF}
\end{eqnarray}\label{examplesDF}%
\end{subequations}
The tensor in Eq.~\eqref{simplestF} is the simplest contribution to $\breve{D}^F$, with $\tilde{D}^{F^+,1}$ and $\tilde{D}^{F^-,1}$, together with LV LECs of order $\mathcal{O}(F_\pi/\Lambda_\chi)$. ($D_{\nu\rho\sigma}^\pm$ contributes to both $\tilde{D}^{F^+,1}$ and $\tilde{D}^{F^+,1}$, due to isospin-breaking from the quark charges.) Eq.~\eqref{loopcorrF} is interesting because this operator gets a contribution from loop corrections due to the dominant $D^-$-dependent $\pi N$ interaction given in Eq.~\eqref{chirallagrDmin}, which is therefore enhanced by a chiral logarithm, cf. Eq.~\eqref{Fplus} below.

\subsection{Heavy-baryon formalism}
Loop calculations in a relativistic meson-nucleon field theory performed with dimensional regularization receive contributions from loop momenta of order $m_N$. Since $m_N/\Lambda_\chi = \mathcal O(1)$, this upsets the assumed power counting. (When more complicated regularization schemes are adopted the power counting can be made consistent, for a review see Ref. \cite{Bernard}.) In heavy-baryon $\chi$PT (HB$\chi$PT) \cite{Jen91}, this problem is overcome by introducing heavy-nucleon fields with fixed velocity $v$, defined by
\begin{eqnarray}\label{Nv}
N_v = \frac{1+\dslash v}{2}e^{i m_N v^\mu x_\mu} N\ ,
\end{eqnarray}
where $p^\mu = m_N v^\mu + k^\mu\ $, 
with $k$ a small residual momentum. Derivatives acting on the heavy fields give, instead of the large nucleon mass, the small residual momenta. Because the propagator of a heavy-nucleon field does not contain the nucleon mass, the results of loop integrals scale with powers of $Q/m_N$ and $Q/\Lambda_\chi$, where $Q$ is of order $m_\pi$ or the external momentum and $\Lambda_\chi \simeq 2\pi \Fp$. In HB$\chi$PT the Dirac matrices are eliminated in favor of the simpler nucleon velocity $v^\mu$ and the covariant spin vector $S^\mu$ with $S=(0,\boldsymbol{\Sigma}/2)$ and $\boldsymbol{\Sigma} = \gamma^5\gamma^0\boldsymbol{\gamma}$ in the nucleon rest frame, where $v=(1,{\boldsymbol 0})$. 

For the LV Lagrangians in Eqs.~\eqref{chirallagrC}, \eqref{chirallagrH}, and \eqref{chirallagrDmin}, we find as leading-order terms in the heavy-baryon formalism%
\begin{eqnarray}
\mathcal{L}_\chi^{\mathrm{HB}} &=& 4\left(\epsilon^{\mu\nu\alpha\beta}\tilde C^{+}_{\rho\alpha\beta}-\tilde{H}_{\mu\nu\rho}\right) v^\rho v_\nu \bar{N}S_\mu N \notag \\
&& + 4 \epsilon^{\mu\nu\alpha\beta}\tilde C^{-}_{\rho\alpha\beta} v_\nu v^\rho \bar{N}\left[\tau_3 - \frac{2}{F_\pi^2 D}\left(\bpi^2\tau_3  - \pi_3 \btau\cdot\bpi\right)\right]S_\mu N\notag \\
&& + \frac{4}{F_\pi D} \epsilon^{\mu\nu\alpha\beta}\tilde{D}^{-}_{\rho\alpha\beta} v^\rho v_\nu \bar{N}(\boldsymbol{\tau}\times\boldsymbol{\pi})_3 S_\mu N \ .
\label{hbchiptLV}
\end{eqnarray}%
All  coupling constants of these interactions scale as $\Lambda_\chi^2/\Lambda_{\rm LV}$ or $\Lambda_\chi \Fp/\Lambda_{\rm LV}$ and thus suffer a suppression of order ${\cal O}(10^{-18,-19})$ compared to LECs appearing in standard $\chi$PT, if $\Lambda_{\rm LV}$ is identified with the Planck scale. In the heavy-baryon limit, the tensors $C^+_{\mu\nu\rho}$ and $H_{\mu\nu\rho}$ lead to an identical leading-order operator. However, because the symmetrization properties of the tensors are different they can, in principle, still be distinguished.

In HB$\chi$PT, the subleading operators in Eqs.~\eqref{chirallagrDplus} give
\begin{eqnarray}
\mathcal{L}_{\chi D^+}^{\mathrm{HB}} &=& 4\breve{D}^+_{\mu\nu\rho\alpha\beta}\epsilon^{\nu\rho\lambda\kappa}v_\lambda v^\alpha v^\beta \bar{N}(\boldsymbol{\tau}\cdot D^\mu \boldsymbol{\pi})S_\kappa N\ ,
\label{subDhbchiptLV}
\end{eqnarray}%
while the terms parametrizing the $D^{\pm}$-dependent nucleon coupling to the photon field in Eq.~\eqref{effDlagr} give
\begin{eqnarray}
\mathcal{L}_{\chi DF}^{\mathrm{HB}} &=& 4e\bar{N}\breve{D}^F_{[\alpha\beta]\mu\nu\rho\sigma\lambda}v^\beta v^\rho v^\sigma v^\lambda S^\alpha N F^{\mu\nu}\ .
\label{subFhbchiptLV}
\end{eqnarray}%
The examples in Eqs.~\eqref{simplestF} and \eqref{loopcorrF} become respectively
\begin{subequations}%
\begin{eqnarray}
\mathcal{L}_{\chi DF}^{\mathrm{HB}} &\ni& 2e\tilde{D}^{F^+,1}_{\mu\nu\rho}\bar{N}S^\mu N F^{\nu\rho} + 2e\tilde{D}^{F^-,1}_{\mu\nu\rho}\bar{N}\tau_3 S^\mu N F^{\nu\rho}\ , \label{simplestFHB} \\
\mathcal{L}_{\chi DF}^{\mathrm{HB}} &\ni& 2e\tilde{D}^{F^+,2}_{\nu[\rho\sigma]}v_\mu v^\nu \bar{N}S^\sigma N F^{\rho\mu} + 2e\tilde{D}^{F^-,2}_{\nu[\rho\sigma]}v_\mu v^\nu \bar{N}\tau_3 S^\sigma N F^{\rho\mu}\ . \label{loopcounterHB}
\end{eqnarray}
\end{subequations}%

The nucleon operators in Eqs.~\eqref{hbchiptLV} and \eqref{subFhbchiptLV} can be used directly as the LV perturbation of the proton or neutron Hamiltonian. As shown in Section~\ref{sec:Expts}, the Hamiltonian can be used to determine LV contributions to observables such as the nucleon spin-precession frequency and transition frequencies in clock-comparison experiments. Taking $v = (1,{\boldsymbol 0})$, we see that, in the nucleon rest-frame, Eq.~\eqref{hbchiptLV} gives exactly the result obtained later on in Eq.~\eqref{LVhamiltonian}. In addition, the heavy-baryon framework greatly simplifies loop calculations, as discussed in the next section. On the other hand, at leading order in the heavy-baryon expansion we neglect terms of order ${\boldsymbol p}/m_N$, such that the results only apply in the ${\boldsymbol p} \rightarrow {\boldsymbol 0}$ limit. Terms of higher order in $\boldsymbol p$, which can become relevant in, for example, storage-ring experiments, can be explicitly calculated in HB$\chi$PT by including subleading terms in the heavy-baryon expansion. However, when such terms are needed below in Sect.~\ref{sec:LVham} we find it more convenient to derive a relativistic expression for the Hamiltonian.

\subsection{Pion-loop diagrams}\label{sec:pionloop}
\begin{figure}
\centering
\includegraphics[scale = 0.7]{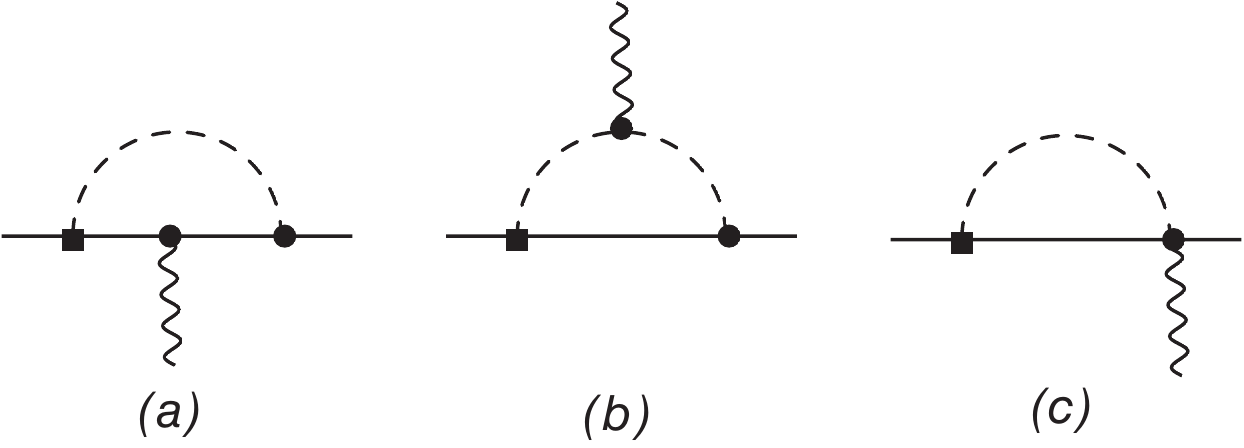}
\caption{Lorentz-violating contributions to the nucleon electromagnetic form factor. The square denotes the pion-nucleon vertex due to the Lorentz-violating tensor $D^-_{\mu\nu\rho}$, while the circles denote leading-order vertices from Eqs.~\eqref{LO1} and \eqref{LO2}. } 
\label{fig:fmdiags}
\end{figure}

In contrast to the $C^{\pm}_{\mu\nu\rho}$ and $H_{\mu\nu\rho}$ components, the LV tensors $D^\pm_{\mu\nu\rho}$ give no contribution to free nucleons at tree level, since we cannot write down a two-point function that does not vanish on-shell. Pion-loop corrections, however, can induce a LV contribution to the electromagnetic form factor via the loop diagrams shown in Fig.~\ref{fig:fmdiags}. The squares represent a LV $\pi N$ vertex from Eq.~\eqref{hbchiptLV}. We assign the external momenta $p$, $p'$, and $q = p-p'$ to the incoming nucleon, the outgoing nucleon, and the photon, respectively. In leading order in the heavy-baryon expansion, we have $v\cdot q = 0$. 

The LV current that follows from the loop calculation has the form
\begin{eqnarray}\label{LVcurrent}
I^\mu(q) &=& i(F^+_{1\,\nu\rho\sigma}(Q^2) + F^-_{1\,\nu\rho\sigma}(Q^2)\tau_3)\,  \epsilon^{\sigma\rho\alpha\beta}v^\nu v_\alpha  \left(Q^2 g^\mu_{\;\;\beta} + q^\mu q_\beta \right)\nonumber\\
&&+(F^+_{2\,\nu\rho\sigma}(Q^2) + F^-_{2\,\nu\rho\sigma}(Q^2)\tau_3)\,v^\nu v^\mu q^{[\sigma} S^{\rho]}\,,
\end{eqnarray}
where $q_{[\sigma} S_{\rho]} = \tfrac{1}{2}(q_\sigma S_\rho - q_\rho S_\sigma)$ and $Q^2 = -q^2$. The loop contributions to the isovector form factors $F^-_{1\,\nu\rho\sigma}(Q^2)$ and $F^-_{2\,\nu\rho\sigma}(Q^2)$ turn out to vanish, while
\begin{eqnarray}
F^+_{1\,\nu\rho\sigma}(Q^2) &=& \tilde{D}^{-}_{\nu\rho\sigma} \frac{e  g_A}{(2\pi F_\pi)^2} \frac{\pi}{3 m_\pi} f_1\left(\frac{Q}{2m_\pi}\right)\,,\nonumber\\
F^+_{2\,\nu\rho\sigma}(Q^2) &=&\tilde{D}^{-}_{\nu\rho\sigma}  \frac{8 e  g_A}{(2\pi F_\pi)^2}  \left[L - \ln \frac{m_\pi^2}{\mu^2} - f_2\left(\frac{Q}{2m_\pi}\right)\right]\,,
\end{eqnarray}
in terms of the two functions%
\begin{subequations}
\begin{eqnarray}
f_1(x) &=& \frac{3}{2x}\left[\frac{x^2+1}{x^2}\arcsin\left(\sqrt{\frac{x^2}{x^2+1}}\right)-\frac{1}{x}\right] \nonumber \\
                & \stackrel{x\ll 1}{=} & 1-\frac{x^2}{5} + \mathcal{O}(x^4)\ , \\
f_2(x) &=& \sqrt{\frac{1+x^2}{x^2}}\ln\left(\frac{\sqrt{1+x^2}+x}{\sqrt{1+x^2}-x}\right)-2 \nonumber \\
                & \stackrel{x \ll 1}{=} & \frac{2x^2}{3} + \mathcal{O}(x^4)\ ,
\end{eqnarray}
\end{subequations}
and $L = 2/(4-d) - \gamma_E + \ln 4\pi$, where $d$ is the number of spacetime dimensions and $\gamma_E$ is the Euler-Mascheroni constant.

The terms proportional to $F^{\pm}_1(Q^2)$ in Eq.~\eqref{LVcurrent} resemble that of the anapole \cite{Zel58} form factor \cite{Mae00}, where the role of the nucleon spin is taken over by a LV absolute direction that depends on $\tilde{D}^-$. Although it is potentially relevant for e.g. electron-nucleon scattering, it does not contribute for on-shell photons and we will neglect this term from now on. 

The terms proportional to $F^{\pm}_2(Q^2)$ do contribute for on-shell photons. In that case, the isoscalar form factor can be written as
\begin{equation}
F^+_{2\,\nu\rho\sigma}(Q^2=0)=  \tilde{D}^{-}_{\nu\rho\sigma}\frac{8 eg_A}{(2\pi F_\pi)^2}\left(L - \ln \frac{m_\pi^2}{\mu^2} \right)\ ,
 \label{Fplus}
\end{equation}
which contains a logarithmic divergence. This divergence needs to be compensated by a counterterm that naturally appears at this order in the chiral expansion, as seen in Eq.~\eqref{loopcounterHB}. The chiral power counting indicates that the long-range contribution from the pion loop and the short-range term in Eq.~\eqref{loopcounterHB} are of similar size. However, the long-range part is somewhat enhanced by the chiral logarithm, as mentioned below Eq.~\eqref{examplesDF}. In any case, a cancellation is unlikely considering the non-analytic dependence of the loop contributions on $m_\pi$ . The isovector piece in Eq.~\eqref{loopcounterHB}, proportional to $\tilde{D}^{F^-,2}_{\nu[\rho\sigma]}$, is not needed for renormalization purposes, but there is no reason to assume it is very small either. 
Absorbing $L$ and the associated $\mu$ dependence into the short-range terms and taking $\mu=m_N$ as the renormalization scale, we obtain for the form factors
\begin{eqnarray}\label{LVformfactors}
F^+_{2\,\nu\rho\sigma}(Q^2=0)&=&  \bar{\tilde{D}}^{F^+,2}_{\nu\rho\sigma}+\tilde{D}^{-}_{\nu\rho\sigma}\frac{8 eg_A}{(2\pi F_\pi)^2}\,\ln \frac{m_N^2}{m_\pi^2} \ ,\nonumber\\
F^-_{2\,\nu\rho\sigma}(Q^2=0)&=&  \tilde{D}^{F^-,2}_{\nu\rho\sigma} \ ,
\end{eqnarray}
where the bar on $\bar{\tilde{D}}^{F^+,2}_{\nu\rho\sigma}$ indicates that this is a renormalized quantity. 

In the following sections, we study the phenomenological consequences of the relativistic LV chiral Lagrangians, obtained in Section~\ref{Npioperators}. The isocalar LV form factor in Eq.~\eqref{LVformfactors} gives a slightly enhanced contribution to the operators summarized in Eq.~\eqref{effDlagr}, which are studied in Section~\ref{sec:EMFF}. We mention already here that, in the rest-frame, the operators that follow from the present loop calculation do not couple to the magnetic field, which is most easily seen from Eq.~\eqref{loopcounterHB}. This is important when considering experimental methods to limit the LV coefficients.

\subsection{Nucleon-nucleon interactions from one-pion exchange}
Our analysis can be extended to systems with multiple interacting nucleons, and in particular to the few-nucleon sector, where $\chi$PT is often called $\chi$EFT. $\chi$EFT allows the derivation of the structure and hierarchy of multi-nucleon interactions (for reviews see e.g. Refs. \cite{Bedaque:2002mn,Epelbaum:2008ga}). $\chi$EFT is usually formulated for nonrelativistic nucleons, which fits naturally with the heavy-baryon framework discussed above. We briefly discuss here the LV $N\!N$ interaction arising from one-pion exchange with the $\pi N$ vertices from Eqs.~\eqref{hbchiptLV} and \eqref{subDhbchiptLV}. Although the tensors $C^\pm$ and $H$ give contributions to $N\!N$ interactions at the same chiral order as $D^+$, we omit them here, because, in contrast to $D^\pm$, there exist nucleon two-point functions for $C^\pm$ and $H$ that already provide very strict limits (see below, in Section~\ref{sec:Expts}).

In combination with the standard leading-order $\pi N$ vertex multiplied by $g_A$, we obtain the LV $N\!N$ potential
\begin{eqnarray}\label{LVpot}
V_{\rm LV} &=& - \left(\epsilon^{ijk}\tilde D^-_{0ij}\right)\frac{2 i g_A}{F_\pi^2} (\btau_1\times\btau_2)_3 \frac{(\bsig_1\cdot{\boldsymbol k})\sigma_2^k +(\bsig_2\cdot{\boldsymbol k})\sigma_1^k}{{\boldsymbol k}^2 + m_\pi^2}\nonumber\\
&&- \left(\epsilon^{jkl}\breve D^+_{ijk00} \right) \frac{4 g_A}{F_\pi} \btau_1\cdot\btau_2 \frac{(\sigma_1^l \sigma_2^m + \sigma_1^m \sigma_2^l)k^i k^m}{{\boldsymbol k}^2 + m_\pi^2}\,\,\,,
\end{eqnarray}
where $\bsig_{1,2}$ $(\btau_{1,2})$ are the spin (isospin) operators of the interacting nucleons and the momentum transfer ${\boldsymbol k} = {\boldsymbol p} - {\boldsymbol p}^{\,\prime}$ flows from nucleon $1$ to nucleon $2$; $\boldsymbol p$ and ${\boldsymbol p}^{\,\prime}$ are the relative momenta of the incoming and outgoing nucleon pair in the center-of-mass frame. The Latin indices $i, j ,k, \dots$ denote spatial directions.  At the same order as the second term in Eq.~\eqref{LVpot}, there exist contributions from LV $N\!N$ contact interactions, which we ignore here.

We postpone a detailed study of this potential and its consequences to future work, but point out that the interactions between nucleons can lead to measurable LV in clock-comparison experiments on nuclei or in the spin precession of nuclei in storage rings. This is especially relevant because the effects could be considerably larger than for nucleons where, in case of the $D^\pm$ operators, a coupling to an electromagnetic field is required. As discussed in Secs.~\ref{spinprec} and \ref{sec:EMFF}, this greatly weakens the constraints on the $D^\pm$ LV tensors. A study of the effects of Eq.~\eqref{LVpot} on, for example, the spin precession of deuterons in storage rings would therefore be very interesting.

\section{Hamiltonian with Lorentz violation}
\label{sec:LVham}
Having obtained the low-energy chiral Lagrangian, we now obtain the limits that are set by existing experimental constraints, from which we deduce which parts of the parameter space have room for improvement. As mentioned, the strictest limits are on the nucleon two-point functions and come from clock-comparison experiments \cite{datatables}. For the analysis of clock-comparison experiments, the block-diagonalized form of the Hamiltonian has proven to be convenient \cite{clockcomparison}. In this diagonal form the Dirac equation for the particle and the antiparticle are decoupled. The diagonalization is achieved by performing a unitary Foldy-Wouthuysen transformation of the fields \cite{Foldwout}. A comparable particle-antiparticle decoupling is obtained in HB$\chi$PT.

The heavy-baryon approach employs a nonrelativistic expansion in $p/M_N$, which implies that observer Lorentz invariance can only be restored perturbatively \cite{rpi}. As for the Foldy-Wouthuysen transformation, for some Hamiltonians an exact diagonalization can be achieved \cite{Foldwout, exactfw}. In most cases, however, the transformation is done such that the off-diagonal parts of the Hamiltonian can be made of arbitrary order in some small quantity. Often, ${\boldsymbol p}^2/m^2$ is chosen as the small parameter, which results in a nonrelativistic expansion of the Hamiltonian, comparable to the heavy-baryon approach. Here, we adopt the approach of Ref.~\cite{kostmuon}, where the relevant Hamiltonian is obtained with a Foldy-Wouthuysen transformation on the relativistic muon Hamiltonian that follows from the mSME \cite{sme}. The small quantities in which the off-diagonal parts of the Hamiltonian are expanded are the LV tensor components and the electromagnetic fields. This results in a relativistic expression for the relevant parts of the Hamiltonian for free nucleons (at least when restricting to frames where the LV coefficients and the EM fields are small with respect to the nucleon mass). 

The Dirac equation that includes the LV from Eqs. \eqref{chirallagrC} and \eqref{chirallagrH} is given by
\begin{equation}
i\partial^0 \psi_w = H_w \psi_w\ ,
\end{equation}
where $w \in \{p,n\}$ denotes proton or neutron and
\begin{eqnarray}
H_w &=&\gamma^0\left(\boldsymbol{\gamma}\cdot\boldsymbol{\Pi} + m_N\right) + eA^0 + \frac{1}{4}(g_w-2)\mu_N \gamma^0 \sigma_{\mu\nu}F^{\mu\nu} \notag \\
&&+ \frac{2}{m_N^2}\tilde{H}_{\alpha\nu\rho}\Pi^\nu \Pi^\rho \Sigma^\alpha - \frac{2}{m_N}\tilde{C}^{w}_{\mu\alpha\beta}\Pi^\mu\gamma^0 \sigma^{\alpha\beta}\ , 
\end{eqnarray}
with $\Pi^\mu = i\partial^\mu - e A^\mu$, $\Sigma^\alpha = \gamma^5\gamma^0\gamma^\alpha$, and where for the proton and neutron $\tilde{C}$ is given by $\tilde{C}^{p} = \tilde{C}^{+} + \tilde{C}^{-}$ and $\tilde{C}^{n} = \tilde{C}^{+} - \tilde{C}^{-}$. We added the term for the anomalous magnetic moment of the nucleon, where $\mu_N = e/2m_N$ is the nuclear magneton.

The operator $H_w$ is not a standard Hamiltonian because it contains extra terms with time derivatives. This is a well-known problem when dealing with LV. In the mSME it can be solved by applying a spinor redefinition that removes the extra time derivatives \cite{Blu97}. However, since we have time-derivative terms of higher order, we use the approach of Ref.~\cite{Kos09}, where one first diagonalizes the Hamiltonian and then substitutes $i\partial^0 \rightarrow \sqrt{{\boldsymbol p}^2 +m^2}$ for the fermion and $i\partial^0 \rightarrow -\sqrt{{\boldsymbol p}^2 +m^2}$ for the antifermion in the LV terms. Contributions that we miss in this way are higher order in the LV components, and hence negligible.

The Foldy-Wouthuysen transformation used to diagonalize $H_w$ is given by $H'_w = e^{\gamma^0\gamma^5\phi}H_w e^{-\gamma^0\gamma^5\phi}$, with $\tan 2\phi=\boldsymbol{\Sigma}\cdot \boldsymbol{\Pi}/m_N$. We assume that all the electromagnetic fields are homogeneous and small and we neglect all contributions that are quadratic in these fields as well as products of LV and an electromagnetic field. This results in a Hamiltonian with off-diagonal $2\times 2$ blocks that are first order in the LV components or the $E$- and $B$-fields. We neglect these small off-diagonal contributions and take the upper left $2\times 2$ block ($h_{w,+}$) as the Hamiltonian for the particle and the lower right $2\times 2$ block ($h_{w,-}$) as the Hamiltonian for the antiparticle. We find that the resulting Hamiltonian is given by
\begin{equation}
h_{w,\pm} = h_{w,0} \pm \delta h_w\ ,
\end{equation}
where $h_{w,0}$ is the conventional particle or antiparticle Hamiltonian, while the LV perturbation $\delta h_w$ is given by
\begin{equation}
\delta h_w = -2\gamma \left[\boldsymbol{\sigma}\cdot\boldsymbol{\bar{\xi}}_w - \gamma \boldsymbol{\sigma}\cdot\boldsymbol{\beta}\left(\bar{\xi}^0_w - \frac{\gamma}{\gamma+1}\boldsymbol{\beta}\cdot\boldsymbol{\bar{\xi}}_w\right) \right]\ ,
\label{LVhamiltonian}
\end{equation}
where $\boldsymbol{\beta} = {\boldsymbol p}/E$ is the (anti)particle velocity, $\gamma$ is the relativistic boost factor, and
\begin{equation}
\bar{\xi}^\mu_w = \xi^{\mu\nu\rho}_w\beta_\nu \beta_\rho = \left[\tilde{H}^{\mu\nu\rho}-\epsilon^{\mu\nu\alpha\beta}(\tilde{C}^{w})^\rho_{\;\;\alpha\beta}\right]\beta_\nu\beta_\rho\ ,
\label{xidefinition}
\end{equation}
with $\beta = (1,\boldsymbol{\beta})$. We thus conclude that the part of $\xi_w^{\mu\nu\rho}$ that is symmetric in $\nu$ and $\rho$ is the only observable combination of the $C$ and $H$ tensors in experiments with free nucleons.
This is consistent with Eq.~\eqref{hbchiptLV}, where the same combination of $C$ and $H$ appears. It confirms that the heavy-baryon and the Foldy-Wouthuysen approach are equivalent for nucleons at rest. The tensor $\xi^{\mu\nu\rho}$ is completely traceless and its observable part therefore has 32 independent components, while the observable parts of $C$ and $H$ both contain 16 independent components. In the following, we will derive bounds on a subset of these.

\section{Experimental constraints}
\label{sec:Expts}
\subsection{Clock-comparison experiments}
The most restrictive limits on Lorentz and CPT violation for protons and neutrons come from clock-comparison experiments \cite{datatables, clockcomparison}. In these experiments transition frequencies of two colocated samples of atoms or ions are compared. The variation of these frequencies, as the clocks rotate with Earth, gives a limit on rotational noninvariance and hence on LV. In Ref.~\cite{clockcomparison} the combinations of mSME tensor components that are observable in clock-comparison experiments are determined by calculating expectation values of the particle Hamiltonian that is linear in LV. For an atom or ion $W$, it is given by
\begin{equation}
h_W' = \sum_w \sum_{N=1}^{N_{w,W}} \delta h_{w,N}\ ,
\end{equation}
where $\delta h_{w,N}$ is the LV Hamiltonian for the $N$th particle of species $w$, the second sum runs over all $N_{w,W}$ particles of species $w$ that are present in the atom or ion $W$, and the first sum runs over all species. In the present case, $\delta h_{w,N}$ for protons and neutrons is given by $\delta h_w$ in Eq.~\eqref{LVhamiltonian}, while it is zero for electrons. 

We take the laboratory $z$ axis as the axis of quantization. The LV corrections to the transition frequencies follow from the expectation value $\delta E(F,M_F) = \left\langle F, M_F | h'_W | F, M_F \right\rangle$, where $\left|F, M_F\right\rangle$ is the state corresponding to the atom or ion with total relevant angular momentum $F$ and projection $M_F$. The LV shift in a frequency corresponding to a transition $(F,M_F) \rightarrow (F',M_F')$ is then given by
\begin{equation}
\delta \omega = \delta E(F,M_F) - \delta E(F',M_F')\ .
\end{equation}
Depending on the rotational transformation properties of the different parts of the Hamiltonian, LV will give rise to different multipole contributions to the transition frequencies. The LV shift can be written as
\begin{equation}
\delta E(F,M_F) = \widetilde{M}^{1}_F E^W_1 + \widetilde{M}^{2}_F E^W_2 + \widetilde{M}^{3}_F E^W_3\ ,
\end{equation}
where the constants $\widetilde{M}^{n}_F$ ($n=1,2,3$), given by ratios of Clebsch-Gordan coefficients, read
\begin{subequations}
\begin{eqnarray}
\widetilde{M}^{1}_F &=& \frac{M_F}{F}\ , \\
\widetilde{M}^{2}_F &=& \frac{3M_F^2 - F(F-1)}{3F^2 - F(F-1)}\ , \\
\widetilde{M}^{3}_F &=& \frac{M_F}{F}\frac{5M_F^2+1-3F(F+1)}{5F^2+1-3F(F+1)}\ .
\end{eqnarray}%
\label{ratioclebschgordan}%
\end{subequations}
Furthermore, $E^W_1$, $E^W_2$, and $E^W_3$ originate from spherical tensors of rank $1,2,$ and $3$, respectively, which require a total angular momentum of at least $\tfrac{1}{2}$, $1$, and $\tfrac{3}{2}$ to be nonvanishing. Following Ref.~\cite{clockcomparison}, we call $E^W_1$, $E^W_2$, and $E^W_3$ the dipole, quadrupole, and octupole contributions, respectively. We calculate these contributions in the nonrelativistic limit, keeping only terms up to first order in ${\boldsymbol p}^2/m_N^2$. We find 
\begin{subequations}
\begin{eqnarray}
E^W_1 &=& \sum_w \left(2\xi_w^{300}\mathcal{M}^{w,W}_1 + (2\xi_w^{300}-\tfrac{3}{5}\xi_w^{(300)})\mathcal{M}^{w,W}_2 - \tfrac{36}{5}\xi_w^{(300)}\mathcal{M}^{w,W}_3\right)\ , \label{dipole} \\
E^W_2 &=& \sum_w 2\epsilon^{3ij}\xi_w^{i(j3)}\mathcal{M}^{w,W}_4\ , \label{quadrupole} \\
E^W_3 &=& \sum_w(\tfrac{3}{5}\xi_w^{(300)} - \xi_w^{333})\mathcal{M}^{w,W}_5\ , \label{octupole}
\end{eqnarray}
\label{Edqo}%
\end{subequations}
where we used that $\xi^{\mu\nu\rho}$ is completely traceless and defined the symmetrized parts of $\xi^{\mu\nu\rho}$ as $\xi^{\mu(\nu\rho)}=\tfrac{1}{2}(\xi^{\mu\nu\rho}+\xi^{\mu\rho\nu})$ and $\xi^{(\mu\nu\rho)}=\tfrac{1}{6}(\xi^{\mu\nu\rho}+\xi^{\mu\rho\nu}+\xi^{\rho\mu\nu}+\xi^{\rho\nu\mu}+\xi^{\nu\mu\rho}+\xi^{\nu\rho\mu})$. In contrast to the mSME case in Ref.~\cite{clockcomparison}, we find an octupole contribution, which originates from the totally-symmetric gluon tensor $H^{\mu\nu\rho}$. Due to the antisymmetry of the $C$ term in Eq.~\eqref{xidefinition}, the contributions of $\xi_w^{(300)}$ contain no component of $H$ or $C$ that is not already present in $\xi_w^{300}$. It does contain different linear combinations of the tensor components, however.

The matrix elements $\mathcal{M}_1$ to $\mathcal{M}_5$ are sums of expectation values of spherical-tensor operators in the special ``stretched'' state $\left|F,F\right\rangle$. The relation between the expectation values in this special state and a state with general $M_F$ follows from the Wigner-Eckart theorem \cite{rose}, which allows to separate the matrix elements of spherical tensors in a Clebsch-Gordan coefficient and a reduced matrix element.
The ratios of these Clebsch-Gordan coefficients for the expectation values in the states $\left|F,F\right\rangle$ and in the states $\left|F,M_F\right\rangle$ are given by the factors $\widetilde{M}^n_F$ in Eqs.~\eqref{ratioclebschgordan}. The relevant expectation values in the state with $M_F = F$ are given by
\begin{subequations}
\begin{eqnarray}
\mathcal{M}^{w,W}_1 &=& -\sum_{N=1}^{N_{w,W}}\left\langle \left[\sigma^3\right]_{w,N}\right\rangle \ , \\
\mathcal{M}^{w,W}_2 &=& \frac{1}{m_N^2}\sum_{N=1}^{N_{w,W}}\left\langle \left[p^3p^j\sigma^j - \sigma^3 p^j p^j\right]_{w,N}\right\rangle \ , \\
\mathcal{M}^{w,W}_3 &=& \frac{1}{m_N^2}\sum_{N=1}^{N_{w,W}}\left\langle \left[p^j p^j \sigma^3\right]_{w,N}\right\rangle \ , \\
\mathcal{M}^{w,W}_4 &=& \frac{1}{m_N^2}\sum_{N=1}^{N_{w,W}}\left\langle \left[p^3(p^1\sigma^2-\sigma^1p^2)\right]_{w,N}\right\rangle \ , \\
\mathcal{M}^{w,W}_5 &=& \frac{1}{m_N^2}\sum_{N=1}^{N_{w,W}}\left\langle \left[5p^3 p^3 \sigma^3 - 2 p^3 p^j \sigma^j - p^j p^j \sigma^3\right]_{w,N}\right\rangle \ .
\end{eqnarray}
\end{subequations}
To determine the values of these matrix elements one has to adopt some nuclear-structure model, such as the simple Schmidt model \cite{Bla52}, wherein the entire angular momentum of the nucleus is carried by a single nucleon. However, we can already see that $\mathcal{M}_2$ to $\mathcal{M}_5$ will most likely be suppressed with respect to $\mathcal{M}_1$, since they contain the small factor ${\boldsymbol p}^2/m_N^2$, which will cause a loss in sensitivity of order ${\cal O}(10^{-2})$.

Finally, we define the physical observable as in Refs.~\cite{clockcomparison} and \cite{spaceclock}, where the transition frequency of a certain clock $A$ is written as $\omega_A = f_A(B_3) + \delta \omega_A$. Here, $f_A(B_3)$ is the conventional transition frequency in terms of the external magnetic field projected on the quantization axis, while $\delta \omega_A$ is the LV contribution. When comparing two clocks, say $A$ and $B$, one defines a frequency $\omega^{\sharp}$ by~\cite{spaceclock}
\begin{subequations}
\begin{equation}
\omega^{\sharp} \equiv \omega_A - f_A(f_B^{-1}(\omega_B)) = \delta \omega_A - \rho \delta \omega_B \ ,
\end{equation}
where
\begin{equation}
\rho=\left.\frac{df_A}{dB_3}\left(\frac{df_B}{dB_3}\right)^{-1}\right|_{B_3=0} \ ,
\end{equation}
\end{subequations}
which in most cases is equal to the ratio of the gyromagnetic ratios of $A$ and $B$. The observable $\omega^\sharp$ vanishes when there is no LV and its explicit form in our case can be obtained by using the expectation values we calculate above. In general it is given by
\begin{eqnarray}
\omega^\sharp = \sum_{n=1}^3 \Delta \widetilde{M}_{F_A}^n E_n^A - \rho \sum_{n=1}^3 \Delta \widetilde{M}_{F_B}^n E_n^B\ ,
\end{eqnarray}
with $\Delta \widetilde{M}_F^n = \widetilde{M}^n_F - \widetilde{M}^n_{F'}$ for a transition $(F,M_F)\rightarrow (F',M_F')$. With the relations in Eqs.~\eqref{Edqo} this can easily be made explicit in terms of $\xi^{\mu\nu\rho}$. To be able to compare different experiments it is convenient to give limits in the Sun-centered inertial reference frame. This frame and the relevant transformations to the laboratory frame are described in Refs.~\cite{spaceclock, datatables}. 

The strongest limits from clock-comparison experiments are on the nonrelativistic dipole contribution to transition frequencies, corresponding to Eq.~\eqref{dipole}. In the mSME the corresponding combination of LV tensor components is called $\tilde{b}_J$, with $J \in \{X,Y,Z\}$ a spatial coordinate in the Sun-centered frame \cite{datatables}. The best bounds on $\tilde{b}_J$ come from a $^3$He/$^{129}$Xe comagnetometer for the $X$ and $Y$ directions \cite{comagnexp, comagntheory} and from a $^{199}$Hg/$^{133}$Cs comagnetometer for the $Z$ direction \cite{hgcs}. These bounds directly translate to a 1$\sigma$ limit on parts of $\xi^{\mu\nu\rho}$, given in the first three rows of Table~\ref{tab:limits}, where the $X$, $Y$, and $Z$ directions are again defined in the Sun-centered frame \cite{datatables}. In obtaining these limits, we have ignored the suppressed contributions proportional to $\mathcal{M}_2$ and $\mathcal{M}_3$ in Eq.~\eqref{dipole}.
\begin{table}[t!]
\centering
\begin{tabular}{ccc}
\hline\hline
 Combination &  Result in GeV &  Ref. \\
\hline
$\left|\xi_n^{XTT}+0.2\xi_p^{XTT}\right|$ & $< 7.3 \times 10^{-34}$ & \cite{comagnexp,comagntheory} \\
$\left|\xi_n^{YTT}+0.2\xi_p^{YTT}\right|$ & $< 7.7 \times 10^{-34}$ & \cite{comagnexp,comagntheory} \\
$\left|\xi_n^{ZTT}+0.1\xi_p^{ZTT}\right|$ & $< 4 \times 10^{-30}$ & \cite{hgcs,hgcsspinratio} \\
$\left|\xi_n^{Y(TX)} + 0.2\xi_p^{Y(TX)}\right|$ & $< 6.9 \times 10^{-28}$ & \cite{neutronboost,comagntheory} \\
$\left|\xi_n^{TTT}+2\xi_n^{X(TX)} + 0.2\left(\xi_p^{TTT}+2\xi_p^{X(TX)}\right)\right|$ & $< 9.0 \times 10^{-28}$ & \cite{neutronboost,comagntheory} \\
$\left|\cos\eta\left(\xi_n^{X(TY)}+0.2\xi_p^{X(TY)}\right)+\sin\eta\left(\xi_n^{X(TZ)}+0.2\xi_p^{X(TZ)}\right)\right|$ & $< 4.0 \times 10^{-28}$ & \cite{neutronboost,comagntheory} \\
$\left|\cos\eta\left(\xi_n^{TTT}+2\xi_n^{Y(TY)}+0.2\left(\xi_p^{TTT}+2\xi_p^{Y(TY)}\right)\right)\right.$ & \multirow{2}{*}{$< 7.4 \times 10^{-28}$} & \multirow{2}{*}{\cite{neutronboost,comagntheory}} \\
$\left.+2\sin\eta\left(\xi_n^{Y(TZ)}+0.2\xi_p^{Y(TZ)}\right)\right| $ &  & \\
\hline\hline
\end{tabular}
\caption{Limits ($1\sigma$) on LV tensor components obtained from two experiments on a $^{3}$He/$^{129}$Xe comagnetometer \cite{comagnexp,neutronboost} and an experiment with $^{199}$Hg/$^{133}$Cs \cite{hgcs}. The contributions of the neutron and the proton correspond to the their relative contributions to the nuclear spin, which are taken from Refs. \cite{heliumspin}, \cite{comagntheory}, and \cite{hgcsspinratio} for $^{3}$He, $^{129}$Xe, and $^{199}$Hg, respectively. The angle $\eta \simeq 23.5^\circ$ is Earth's axial tilt.}
\label{tab:limits}%
\end{table}

Because of the high sensitivity of the clock-comparison experiments, also boost effects due to Earth's velocity can be used to bound LV parameters. These effects are suppressed by at least one power of the velocity of Earth with respect to the Sun, $\beta_\oplus \simeq 10^{-4}$. The dominant signal will oscillate with the rotation frequency of Earth around the Sun. This can be seen by realizing that the transformation from the laboratory frame to the Sun-centered frame, to first order in $\beta_\oplus$, is given by \cite{spaceclock}
\begin{equation}
\Lambda = \left(\begin{array}{>{\centering\arraybackslash$} p{1.5em} <{$} >{\centering\arraybackslash$} p{1.5em} <{$}}
1 & {\bf 0} \\
{\bf 0}^T & \mathcal{R} 
\end{array}\right)
\cdot
\left(\begin{array}{>{\centering\arraybackslash$} p{1.5em} <{$} >{\centering\arraybackslash$} p{1.5em} <{$}}
1 & \boldsymbol{\beta}_\oplus \\
\boldsymbol{\beta}_\oplus^T & {\bf 1} 
\end{array}\right)\ ,
\label{frametransformation}
\end{equation}
where $\mathcal{R}$ is the rotation matrix that rotates the axes of the instantaneous rest-frame of the laboratory into the axes of the Sun-centered frame. We neglect the rotation velocity of the Earth around its axis, which is almost two orders of magnitude smaller than its orbital velocity. The entries of the first matrix in Eq.~\eqref{frametransformation} are of order 1, while the off-diagonal entries of the second transformation are of order $10^{-4}$, which demonstrates the suppression of boost effects. 

An analysis looking for a boost-dependent signal oscillating with the rotation frequency of Earth around the Sun was performed in Ref.~\cite{neutronboost}. The last four rows in Table~\ref{tab:limits} represent this result in terms of $\xi^{\mu\nu\rho}$, where we used Ref.~\cite{comagntheory} to obtain the sensitivity to the proton parameters.

Of the 32 observable LV components of $\xi^{\mu\nu\rho}$, i.e. the traceless part of $\xi^{\mu(\nu\rho)}$, only 7 combinations are bounded by the limits in Table~\ref{tab:limits}. However, if we ignore the possibility of accidental cancellations among different LV tensor components, Table~\ref{tab:limits} represent bounds on 10 independent components of $\xi^{\mu\nu\rho}$. If we include, from Eq.~\eqref{dipole}, the corrections proportional to $\mathcal{M}_2$ and $\mathcal{M}_3$, ten additional $\xi$ components will receive limits that are weaker by a factor ${\boldsymbol p}^2/m_N^2 \simeq 10^{-2}$. These are limits on the tensor components whose indices are a permutation of the ones present in the Table. Taking into account the tracelessness of $\xi^{\mu\nu\rho}$, we conclude that every component of $\xi^{\mu(\nu\rho)}$ that has at least one time-like index has a bound between $10^{-25}$ and $10^{-33}$ GeV for the neutron, while the sensitivity to the proton tensor is a factor 5 to 10 worse. 

To obtain bounds on the remaining 12 components, which all have only space indices, one should consider double-boost effects, which are suppressed by $\beta_{\oplus}^2 \simeq 10^{-8}$ with respect to the dominant effects. For definite limits, a dedicated analysis of the data would be necessary, since the signal will contain higher harmonics of Earth's orbital frequency. We estimate that components of $\xi^{\mu(\nu\rho)}$ that only have space indices will receive bounds between $10^{-21}$ and $10^{-25}$ GeV for the neutron, while proton bounds will again be 5 to 10 times less stringent.

By using the definition of $\xi^{\mu\nu\rho}$ in Eq.~\eqref{xidefinition} and those of $\tilde{C}^{\pm}$ and $\tilde{H}$, one can easily translate the bounds in Table~\ref{tab:limits} to a (less compact) form  that explicitly shows the original quark and gluon parameters. For example, the $\xi$ component with the best bound becomes
\begin{eqnarray}
(\xi_n)^{XTT} + 0.2(\xi_p)^{XTT} & = & 1.2 h H^{TTX} + 2(0.6c^+ - 0.4c^-)(C^u)^{T[YZ]} \nonumber \\
                                                          &&      + 2(0.6c^+ + 0.4 c^-)(C^d)^{T[YZ]} \ .
\end{eqnarray}
The NDA estimates for the LECs are $h=\mathcal{O}(\Lambda_\chi^2)$ and $c^\pm = \mathcal{O}(\Lambda_\chi F_\pi)$. Assuming that the combinations $0.6c^+ \pm 0.4c^-$ are of the same order as $c^\pm$ itself and ignoring the possibility of accidental cancellations, we see that this clock-comparison observable gives a limit on $H$ and $C^q$ of order $10^{-33}$ GeV$^{-1}$. Repeating this analysis for all results in Table~\ref{tab:limits}, gives us the order-of-magnitude bounds in Table~\ref{tab:sens}.  

As mentioned below Eqs.~\eqref{Edqo}, the corrections proportional to $\mathcal{M}_2$ and $\mathcal{M}_3$ do not contain any new components of $H$ or $C^q$. We thus conclude that clock-comparison experiments allow us to place limits on 9 of the 16 components of $H_{\mu\nu\rho}$, and on 13 of the 16 components of $C^q_{\mu[\nu\rho]}$. The remaining components of $H_{\mu\nu\rho}$ have only space indices, while those of $C^q_{\mu\nu\rho}$ are $C^q_{Y[XY]}$, $C^q_{X[YX]}$, and all components of the form $C^q_{J[TK]}$, with $J,K \in \{X,Y,Z\}$. For these, double-boost effects should be able to provide bounds of order $10^{-21}$ to $10^{-25}$ GeV$^{-1}$.

We find that clock-comparison experiments bound components of $H$ and $C^q$ at least two orders beyond the Planck scale. However, to reach this conclusion, we assumed that no cancellations between LV tensor components take place. When one allows for accidental cancellations up to 1\%, one concludes that it is desirable to get a few orders of magnitude improvement for the worst-bounded components of $\xi$, $C^q$, or $H$. Such improvements can be provided by spin-precession experiments and in particular by storage-ring experiments, since the latter do not suffer from boost suppressions.
\begin{table}[t!]
\centering
\begin{tabular}{cc}
\hline\hline
 Tensor component &  Limit in GeV$^{-1}$ \\
\hline
$H_{TTX}$, $H_{TTY}$, $C^q_{T[XZ]}$, $C^q_{T[YZ]}$ & $<10^{-33}$ \\
$H_{TTZ}$ & $<10^{-30}$ \\
$C^q_{T[XY]}$ & $<10^{-29}$ \\
$H_{TXX}$, $H_{TXY}$, $H_{TYY}$ & $<10^{-28}$ \\
$H_{TTT}$, $H_{TXZ}$, $H_{TYZ}$, $C^q_{T[TZ]}$, $C^q_{X[YZ]}$, $C^q_{Y[XZ]}$, $C^q_{Y[YZ]}$ & $<10^{-27}$ \\
$C^q_{T[TX]}$, $C^q_{T[TY]}$, $C^q_{T[TZ]}$, $C^q_{X[XZ]}$, $C^q_{Z[XZ]}$, $C^q_{Z[YZ]}$ & $<10^{-26}$ \\
\hline\hline
\end{tabular}
\caption{Order-of-magnitude bounds on the LV tensor components defined in Eqs.~\eqref{lagr1gevQ} and \eqref{lagr1gevG}. We apply a logarithmic way of rounding: a factor larger than $10^{0.5}$ is rounded to $10$, while anything smaller than $10^{0.5}$ is rounded to $1$. Since we do not know the size of the LECs $c^\pm$, we take the limits on $C^u$ and $C^d$ to be the same and summarize them as $C^q$.}
\label{tab:sens}%
\end{table}

\subsection{Spin-precession experiments}\label{spinprec}
A different class of experiments that can be used to bound LV for nucleons and light nuclei are experiments that measure their spin-precession frequency. In many experiments, the ratio of the spin-precession frequency ($\omega_s$) to the cyclotron frequency ($\omega_c$) is measured, which can be used to determine the $g$ factor.
The dominant LV contribution to the nucleon cyclotron frequency turns out to be proportional to the electromagnetic field times a LV tensor component. This results in a suppression by a factor of $\sim 10^{-15}$ with respect to the LV effect on the spin-precession frequency, which is first order in LV. We therefore neglect this LV contribution and take the cyclotron frequency as conventional.

The spin-precession frequency follows from the Heisenberg equation of motion for the spin operator,
\begin{equation}
 \frac{d\boldsymbol{\sigma}}{dt} = -i[\boldsymbol{\sigma},H] = \boldsymbol{\omega_s}\times \boldsymbol{\sigma} \ .
 \end{equation}
 By using the Hamiltonian from Eq.~\eqref{LVhamiltonian} this gives a LV contribution to the spin-precession frequency of nucleon $w$ of
\begin{equation}
\mp\frac{\delta\boldsymbol{\omega}_{s,w}}{4} = -\gamma \boldsymbol{\bar{\xi}}_w + \gamma^2 \boldsymbol{\beta}\left(\bar{\xi}^0_w -\frac{\gamma}{\gamma+1} \boldsymbol{\beta}\cdot\boldsymbol{\bar{\xi}}_w\right) ,
\label{generalspinprec}
\end{equation}
where the upper (lower) sign applies to particles (antiparticles). 
If we assume the particle performs an integer number of revolutions in an experiment in a magnetic storage ring, where there is no $E$ field and $\boldsymbol{\beta}\cdot {\bf B} = 0$, we obtain that the LV correction to the absolute value of the average spin-precession frequency is given by
\begin{equation}
|\left\langle \delta \omega_{s,w} \right\rangle| = \pm 2\gamma\left(2\xi_w^{k00} + \boldsymbol{\beta}^2(\xi_w^{k00}-\xi_w^{klm}\hat{B}^l\hat{B}^m)\right)\hat{B}^k\ .
\label{omegasLV}
\end{equation}

We can apply Eq.~\eqref{omegasLV} to a comparison of the $g$ factors of the proton \cite{protong} and the antiproton \cite{antiprotong}. Both of these are determined by measuring $\omega_s/\omega_c$ in a double Penning trap. Neglecting the velocity-dependent part of Eq.~\eqref{omegasLV}, we obtain for the experiments in Refs.~\cite{protong,antiprotong} that
\begin{equation}
\frac{|\left\langle\omega_{s,p}\right\rangle|}{\omega_{c,p}} - \frac{|\left\langle\omega_{s,\bar{p}}\right\rangle|}{\omega_{c,\bar{p}}} = \frac{4}{\omega_{c,p}}\left[\xi_p^{100}\right]_{\mathrm{Mainz}} + \frac{4}{\omega_{c,\bar{p}}}\left[\xi_p^{300}\right]_{\mathrm{CERN}} = (2.4 \pm 12)\times 10^{-6}\ ,
\end{equation}
where the subscripts ``CERN'' and ``Mainz'' denote that the LV tensors are defined in the laboratory frame of the corresponding experiments, with the $\hat{x}$, $\hat{y}$, and $\hat{z}$ axes pointing south, east, and up, respectively. We took the magnetic field in the proton (antiproton) experiment as pointing south (up). Averaging over one full sidereal day, we obtain that in the Sun-centered frame \cite{datatables}
\begin{eqnarray}
\frac{|\left\langle\omega_{s,p}\right\rangle|}{\omega_{c,p}} - \frac{|\left\langle\omega_{s,\bar{p}}\right\rangle|}{\omega_{c,\bar{p}}} = 4\xi_p^{ZTT}\left(\frac{\cos\zeta_{\mathrm{CERN}}}{\omega_{c,\bar{p}}}-\frac{\sin\zeta_{\mathrm{Mainz}}}{\omega_{c,p}}\right)\ ,
\end{eqnarray}
where the colatitudes of CERN and Mainz are given by $\zeta_{\mathrm{CERN}} = 43.8^\circ$ and $\zeta_{\mathrm{Mainz}} = 40.0^\circ$, respectively.
This translates to a (1$\sigma$) limit of
\begin{equation}
|\xi_p^{ZTT}| < 2.7 \times 10^{-21}\ \mathrm{GeV}\ .
\label{ratiobound}
\end{equation}
While this does not provide the same sensitivity as clock-comparison experiments,  it is a clean and direct limit on the proton tensor components, independent of a nuclear model.

A similar result can be obtained from storage-ring experiments~\cite{jedi,Ana15}. An important difference with the Penning-trap experiments is that there will be a significant contribution from the boost-dependent part of Eq.~\eqref{omegasLV}, with sensitivity to $\xi_p^{J(KL)}$, for which we were unable to derive definite bounds from clock-comparison experiments. For example, in Ref.~\cite{jedi} an absolute precision on the spin tune, defined by $\nu_s = |\omega_s|/|\omega_c|$, of  $\sigma_{\nu_s} \simeq 3\times 10^{-8}$ per year is claimed. Analyzing the sidereal variation of the spin tune would then give access to, for example, $\xi_p^{Z(ZJ)}$. When it becomes possible to store polarized antiprotons in the same ring, a ratio of the $g$ factors of the proton and antiproton can be measured with a precision of $10^{-9}$ or better. This is about three orders of magnitude better than the result used to obtain Eq.~\eqref{ratiobound} and also has sensitivity to the boost-dependent part of Eq.~\eqref{omegasLV}. Hence, storage-ring experiments can provide improved or complementary bounds on $\xi^{\mu(\nu\rho)}$ by analyzing the sidereal variation of the spin tune or by comparing the spin tune of particles and antiparticles. In addition to giving access to boost-dependent parts of the observables, they give more direct bounds that do not depend on nuclear models.

\subsection{Electromagnetic form factor}\label{sec:EMFF}
From the Lagrangian density in Eq.~\eqref{effDlagr}, we can construct a Hamiltonian and do a Foldy-Wouthuysen transformation, just as we did to obtain the Hamiltonian in Eq.~\eqref{LVhamiltonian}. In this case we get
\begin{equation}
\delta h_w = -2\gamma\boldsymbol{\sigma}\cdot\left[\boldsymbol{\Upsilon}_w - \frac{\gamma}{\gamma+1}\boldsymbol{\beta} \Upsilon^0_w\right]\ ,
\label{Dhamiltonian}
\end{equation}
with
\begin{equation}
\Upsilon_w^{\alpha} = -2\gamma^2 eF_{\mu\nu}(\breve{D}^{w})^{[\alpha\beta]\mu\nu\rho\sigma\lambda}\beta_\beta\beta_\rho\beta_\sigma\beta_\lambda\ ,
\end{equation}
where $\breve{D}^{p}$ ($\breve{D}^{n}$) is the upper-left (lower-right) entry of the isospin matrix $\breve{D}$ in Eq.~\eqref{effDlagr}. By using Eq.~\eqref{Dhamiltonian}, we can derive the LV correction to transition frequencies in clock-comparison experiments and to the nucleon spin-precession frequency. The latter is given by
\begin{equation}
\mp\frac{\delta\boldsymbol{\omega}_{s,w}}{4} = -\gamma \boldsymbol{\Upsilon}_w + \frac{\gamma^2}{\gamma+1} \boldsymbol{\beta}\Upsilon^0_w \ .
\label{generalspinprecups}
\end{equation}
Bounds can again be obtained from clock-comparison experiments or from Penning-trap and storage-ring experiments. However, these bounds will be significantly weaker than those in Table~\ref{tab:limits}, because of the presence of the electromagnetic field strength in $\Upsilon^\mu$. 

Focussing on the operator described by Eq.~\eqref{simplestF}, for example, we find
\begin{subequations}
\begin{eqnarray}
\Upsilon_p^{\alpha} &=& -\frac{e}{\gamma^2}(g^{\alpha\beta}-\gamma^2\beta^\alpha \beta^\beta) (\tilde{D}^{F^+,1}+\tilde{D}^{F^-,1})_{\beta\mu\nu}F^{\mu\nu}\ , \\
\Upsilon_n^{\alpha} &=& -\frac{e}{\gamma^2}(g^{\alpha\beta}-\gamma^2\beta^\alpha \beta^\beta) (\tilde{D}^{F^+,1}-\tilde{D}^{F^-,1})_{\beta\mu\nu}F^{\mu\nu}\ .
\end{eqnarray}
\end{subequations}
Taking the nonrelativistic limit (and neglecting $E$ fields), we see that the LV component that replaces $\xi_w^{300}$ in the first term of the dipole contribution in Eq.~\eqref{dipole} is
\begin{equation}
\left. \Upsilon_p^{3}\right|_{\rm NR\ limit} = eB^l \epsilon^{jkl}(\tilde{D}^{F^+,1}+\tilde{D}^{F^-,1})^{3jk}\ ,
\label{upsdipole}%
\end{equation}
with the opposite sign for $\tilde{D}^{F^-,1}$ in the neutron case. Here, $B^l$ is the $l$-component of the magnetic field (its direction differed in the different runs of the experiment in Ref.~\cite{comagnexp}). In principle we could continue and obtain bounds on the corresponding components of $D^{\pm}_{\mu\nu\rho}$. Unfortunately, in contrast to $\xi_w^{300}$, Eq.~\eqref{upsdipole}, when transformed to the Sun-centered frame, induces terms that oscillate with twice the sidereal frequency, making a reanalysis of the pertinent data necessary to obtain exact limits. 

It is also important to mention that some operators, like the loop-induced operators obtained in Section~\ref{sec:pionloop}, do not couple to the magnetic field in the rest-frame of the nucleon. This implies that bounds from clock-comparison experiments for such operators will suffer another loss of ${\boldsymbol p}^2/m_N^2 \simeq 10^{-2}$. Next to giving more direct limits, storage-ring experiments can have an advantage in such cases, because boost effects are not suppressed.

In general, LV effects in the mentioned experiments due to $D$ will be suppressed by a factor $e F^{\mu\nu}F_\pi/\Lambda_\chi^3$ or $e F^{\mu\nu}/\Lambda_\chi^2$ with respect to effects induced by $C$ or $H$. For a magnetic field of 1 Tesla this constitutes a suppression factor of order ${\cal O}(10^{-16})$. From clock-comparison experiments we can thus at best expect bounds of order ${\cal O}(10^{-17})$ GeV$^{-1}$ on $D^{\pm}_{\mu\nu\rho}$, while some other, double-boost dependent, components will receive bounds of order ${\cal O}(10^{-9,-6})$ GeV$^{-1}$. None of the bounds on $D$ that can be obtained from clock-comparison experiments are thus at Planck-scale level.

\section{Summary and outlook}
\label{sec:Summary}
In this paper, we extended chiral perturbation theory, the effective field theory of QCD for nucleons and pions at low energy, with interactions that violate Lorentz and CPT invariance. In our exploratory study, we took the dominant operators that result from dimension-5 Lorentz- and CPT-violating operators with quark, gluon, and photon fields. We studied two quark-gluon interaction terms and one pure-gluon term. The LV arising from these terms is parametrized by the tensor components $C^\pm$, $D^\pm$, and $H$. We derived the dominant chiral Lagrangian arising from the corresponding LV quark-gluon terms and its heavy-baryon limit.
We calculated several pion-loop diagrams that are relevant to the nucleon electromagnetic form factor and derived the relativistic LV Hamiltonian.

The symmetries dictate that the dominant contributions for $C^\pm$ and $H$ are nucleon two-point functions, while for $D^\pm$ no nonvanishing two-point function exists. This results in far better bounds on $C^\pm$ and $H$ than on $D^\pm$, due to the in standard $\chi$PT unfamiliar feature that the strictest limits arise from two-point functions, through the frame-dependent observables that they induce. The limits on two-point functions were obtained from experiments on clock comparisons with nuclei and on the spin rotation of nucleons in penning traps. Compared to the bounds obtained in Table I of Ref.~\cite{dim5lagrangian}, our best bounds are about four orders of magnitude better. This results mainly from the use of updated experimental results. We concluded that bounds on $C^\pm$ and $H$ are at or beyond Planck-scale level, although a few orders improvement would be desirable for some of the components. Such improvements could be provided by storage-ring experiments. 

For the $D^\pm$ coefficients we derived the contribution to the nucleon electromagnetic form factor and to $N\!N$ interactions from one-pion exchange. Using the nucleon electromagnetic form factor, we estimated that potential limits on the $D^\pm$ coefficients, considering only one-nucleon effects, are still several orders of magnitude from the Planck scale. Also here storage rings could make major contributions, although the Planck scale is likely to stay out of reach for nucleons. However, the contribution of $D^\pm$ to $N\!N$ interactions in nuclei could provide much better bounds. Especially the spin precession of the deuteron in storage rings is promising in this respect.

Our research could be extended in several directions. The complete set of LV quark and gluon operators should be studied and the potential of other experimental observables should be explored. We addressed the spin precession of nucleons in magnetic storage rings. Definite plans~\cite{jedi,Ana15} exist to search for electric dipole moments of the proton and the deuteron in this way. Such experiments can be adapted to search for Lorentz and CPT violation as well. As mentioned above, compared to $C^{\pm}$ and $H$, $D^{\pm}$-related effects in nucleons are suppressed by a factor $1/\Lambda^2_\chi$ and the occurrence of $E$ or $B$ fields, which reduces the sensitivity by orders of magnitude. It would therefore be interesting to extend our study of LV in $N\!N$ interactions to the deuteron. Because of its simple structure, the deuteron would be particularly interesting to study, along the lines of $\chi$PT analyses of its P-odd~\cite{Kap99,Sav01} and T-odd~\cite{Vri11,Mer13} electromagnetic form factors. We expect that in this way the limits on $D^\pm$ can be improved by many orders of magnitude.

\section{Acknowledgments}
We thank H. Wilschut, M. Schindler, and B. Altschul for helpful suggestions.
This research was supported by the Dutch foundation for fundamental research FOM under programs 
104 and 156 and by the Dutch Organization for Scientific Research (NWO) through a VENI grant (J. de V.).
J. N. acknowledges the financial support of the Portuguese Foundation for Science and Technology (FCT)
under grant SFRH/BPD/101403/2014 and program POPH/FSE.

\appendix

\section{Construction of the chiral Lagrangian}
\label{app:so4}
\subsection{SO(4) formalism}
We briefly summarize the techniques for the construction of the chiral Lagrangian in the $SO(4)$ formalism of $\chi$PT and refer to Ref.~\cite{Wei96} for more details. We focus on QCD with two flavors, which is approximately globally invariant under $SU(2)_L\times SU(2)_R$ transformations of the quark doublet $Q=(u\,d)^T$,
\begin{equation}
Q \rightarrow Q' =  \exp\left[i {\boldsymbol\theta}_V\cdot {\boldsymbol t} + i {\boldsymbol\theta}_A \cdot {\boldsymbol t}\,\gamma_5 \right]\,Q\,\,\,,
\end{equation}
where ${\boldsymbol\theta}_{V,A}$ are real vectors and $t_a = \tau_a/2$, where $\tau_a$ are the Pauli isospin matrices. The chiral group is isomorphic to the group of $SO(4)$ rotations in Euclidean space. This global $SO(4)$ symmetry is spontaneously broken to its $SO(3)$ isospin subgroup. The Goldstone bosons, the pions, live in the coset space $SO(4)/SO(3)$ also known as the ``chiral circle''. It is convenient to parametrize this chiral circle in terms of dimensionless fields $\bzeta = \bpi/F_\pi$, where $\bpi$ is the pion field, and to introduce the orthogonal $4\times 4$ rotation matrix
\begin{eqnarray}\label{ChiRotation}
R_{\alpha\beta}=\left(\begin{array}{cc}\delta_{ij}
-\frac{2}{D}\zeta_i \zeta_j&\frac{2}{D}\zeta_i \\
-\frac{2}{D}\zeta_j &\frac{1}{D}
\left(1-\zeta^{\,2}\right)\\
\end{array}\right),
\end{eqnarray}
where $D = 1+ \zeta^{\,2}$.

The field $\bzeta$ transforms as a vector under isospin transformations,
\begin{equation}
\delta \bzeta = \boldsymbol{\theta}_V \times\bzeta\ ,
\end{equation}
but it transforms nonlinearly under axial transformations,
\begin{equation}
\delta\bzeta = (1-\zeta^{\,2})  \boldsymbol{\theta}_A + 2 (\boldsymbol{\theta}_A \cdot\bzeta)\,\bzeta\ . 
\end{equation}
For this reason it is convenient to introduce a chiral covariant derivative $D_\mu\bzeta = (\partial _\mu \bzeta)/D$, which transforms as
\begin{equation}
\delta (D_\mu \bzeta) =2( \bzeta \times \boldsymbol{\theta}_A)\times D_\mu \bzeta\, ,
\end{equation}
such that $(D_\mu \bzeta\,)^2$, the pion kinetic energy term in Eq.~\eqref{LO1}, is invariant under isospin and axial transformations. Similarly, we can introduce a nucleon doublet $N$ that we define to transform as the pion covariant derivative, but in the isospin-$1/2$ representation, that is
\begin{equation}
\delta N = i (\boldsymbol{t}\cdot\boldsymbol{\theta}_V) \,N + 2i\ \boldsymbol{t}\cdot(\bzeta\times\boldsymbol{\theta}_A)\, N\,.
\end{equation}
Because the axial transformation is spacetime dependent (it contains $\bzeta$), the derivative $\partial_\mu N$ transforms different from $N$ itself.  To remedy this we also  introduce the nucleon chiral covariant derivative 
\begin{equation}
\mathcal D_\mu N = (\partial_\mu + 2i\ \boldsymbol{t}\cdot \bzeta \times D_\mu \bzeta)\,N\ ,
\end{equation}
which does transform as $N$ itself. 

The chiral Lagrangian that corresponds to the chiral-invariant part of the QCD Lagrangian can now be obtained by constructing all operators consisting of $D_\mu \bzeta$, $N$, and $\mathcal D_\mu N$ that transform in the same way under $C$, $P$, $T$, and Lorentz transformations as the corresponding terms in the QCD Lagrangian. This then gives rise to the terms in Eq.~\eqref{LO1}, which are Lorentz invariant, and Eqs.~\eqref{chirallagrCplus}, \eqref{chirallagrH}, and \eqref{chirallagrDplus}, which violate Lorentz symmetry.  

The formalism to include chiral-symmetry-breaking operators in the $SU(2)_L\times SU(2)_R$ $\chi$PT Lagrangian is outlined in Ref.~\cite{Wei96}. Operators that break the symmetry as components of chiral tensors can be obtained by rotating operators constructed with non-Goldstone fields $\Psi$,
such as nucleons and nucleon and pion covariant derivatives,
\begin{eqnarray}\label{ChiBreak}
\mathcal O_{ij\cdots z}[\bzeta, \Psi] 
= R_{i \alpha} R_{j \beta} \cdots R_{z \xi} \mathcal O_{\alpha\beta\cdots \xi}[0, \Psi]\ .
\end{eqnarray}
Chiral-symmetry-breaking terms in the QCD Lagrangian induce effective interactions that contain $\bzeta$ directly, without derivatives. As an example, we consider the quark-mass term $\bar m\,\bar Q Q$ which transforms as the fourth component of an $SO(4)$ four-vector. 
From Eqs. \eqref{ChiBreak} and \eqref{ChiRotation} we obtain
\begin{eqnarray}
S_4[\pi, \Psi] &=& \frac{1}{D}
\left(1-\zeta^{\, 2}\right) S_4[0, \Psi] - \frac{2 \bzeta}{ D}\cdot {\boldsymbol S}[0, \Psi]\ .
\end{eqnarray}
Since the quark mass is a Lorentz scalar and the pion field a Lorentz pseudoscalar, $S_4[0, \Psi]$ has to be even under $P$ and $T$ transformations, while ${\boldsymbol S}[0, \Psi]$ has to be $P$- and $T$-odd. A choice is then $S[0,0] = \bar m(\mbox{\boldmath $0$},\; v_0)$, where $v_0$ is a real number that depends on the details of the spontaneous breaking of chiral symmetry. This choice generates
\begin{eqnarray}\label{pionmass}
S_4[\pi,0]= \frac{1}{D}\left(1-\zeta^{\, 2}\right)\bar m v_0 = \bar m v_0 - \bar m v_0 \frac{2\zeta^{\,2}}{D}\ .
\end{eqnarray}
The first term in Eq. \eqref{pionmass} is an irrelevant constant, while the second term give the first contribution to the pion mass in Eq.~\eqref{LO2}, after the identification $m_\pi^2 = 4 v_0 \bar m/F_\pi^2$. 

In exactly the same way we obtain the operators in Eqs.~\eqref{chirallagrCmin} and \eqref{chirallagrDmin} by using that the $C^-$ and $D^-$ LV tensors transform as components of the $SO(4)$ tensor in Eq.~\eqref{eq:tensors}. As an example we consider the $C^-_{\mu\nu\rho}$ operator in Eq.~\eqref{chirallagrCmin}. It transforms as the 34-component of the $SO(4)$ tensor given in Eq.~\eqref{eq:tensors}. In this case we obtain from Eqs. \eqref{ChiBreak} and \eqref{ChiRotation}
\begin{eqnarray}
T_{34}[\pi, \Psi] &=& \left[\left(1-\frac{2\zeta^{\,2}}{D}\right)\delta_{i3} + \frac{2\zeta_3\zeta_i}{D}\right]T_{i4}[0, \Psi] + \frac{2}{D}\left[\zeta_i\delta_{j3} - \zeta_j\delta_{i3}\right]T_{ij}[0, \Psi]\ .
\end{eqnarray}
The candidate operators are restricted by the $C$, $P$, and $T$ properties given in Table~\ref{tab:cptproperties}. This leaves as a possible choice the tensor $T_{ij}[0,\Psi] = T_{44}[0,\Psi] = 0$ and $T_{i4}[0,\Psi] = -T_{4i}[0,\Psi] = \frac{ic^-}{m_N}C^-_{\mu\nu\rho}\bar{N}\tau_i \sigma^{\nu\rho}\mathcal{D}^\mu N$, which gives rise to the operator in Eq.~\eqref{chirallagrCmin}.

\subsection{Naive dimensional analysis}
The procedure described above allows for the construction of the chiral Lagrangian order by order in the chiral expansion. However, it does not predict the sizes of the LECs associated with each interaction. Nevertheless, the chiral expansion relies on the LECs to have a size within a certain natural range in order to have a consistent chiral power counting. The sizes of the LECs can then be estimated by naive dimensional analysis (NDA) \cite{Man84}. Ideally, the LECs are fitted to data or calculated with nonperturbative methods such as lattice QCD, but for the study of LV interactions neither data nor lattice calculations exist. We therefore rely on NDA to give an order-of-magnitude estimate of the LECs, but we stress that these estimates are associated with  a significant uncertainty.

The NDA rules can be neatly summarized by introducing the concept of a reduced coupling \cite{Wei89NDA}. Consider an interaction term with dimension $D$ and $N$ fields and coupling constant $g$. The reduced coupling is defined as
\begin{eqnarray}
g^R = \Lambda^{D-4}(4\pi)^{2-N} g,
\end{eqnarray}
where $\Lambda$ is the scale at which two theories are matched (the $\chi$PT and QCD Lagrangians), in our case identified with $\Lambda_\chi \sim 2\pi F_\pi \sim m_N$. The NDA estimate of a LEC appearing in the chiral Lagrangian is obtained by  demanding that the reduced coupling
of an operator below $\Lambda_\chi$ is of the same size as the product of the reduced couplings of
the operators that appear above $\Lambda_\chi$ and induce the low-energy interaction. For instance, consider the contribution of the quark masses to the pion mass. The reduced pion mass is given by $(m_\pi^2)^R = m_\pi^2/\Lambda_\chi^2$, whereas the reduced quark mass is given by $(\bar m)^R = \bar m/\Lambda_\chi$. The NDA rule then gives $ m_\pi^2 = \mathcal O(\bar m \Lambda_\chi)$, which agrees fairly well with the actual pion mass. 
In full QCD we could have dressed the quark mass by any number of gluon interactions, which would bring in factors of $g_s^R = g_s/(4\pi)$. Consistency of the estimates then requires that we count $g_s \sim 4 \pi$. Another example would be the coupling constant of the standard CPT-even axial-vector pion-nucleon coupling in Eq.~\eqref{LO1} with reduced coupling constant 
$(g_A/F_\pi)^R =(g_A/F_\pi)\Lambda_\chi/(4\pi) \simeq g_A$. This interaction arises from the chiral-invariant part of the QCD Lagrangian where the reduced couplings are $\mathcal O(1)$, because we count $g_s \sim 4\pi$. This implies that $g_A = \mathcal O(1)$, consistent with the actual value $g_A\simeq 1.27$. 

We now apply the NDA procedure to estimate some of the LECs appearing in the LV chiral Lagrangian. The terms in Eq.~\eqref{chirallagrC} have reduced couplings $(\tilde C^\pm_{\mu\nu\rho}/m_N)^R = \tilde C^\pm_{\mu\nu\rho}/m_N$, which should be equal to the reduced coupling of the LV quark-gluon interaction, $(C^\pm_{\mu\nu\rho})^R=C^\pm_{\mu\nu\rho} (\Lambda_\chi/4\pi)$. The NDA rule then estimates $\tilde C^\pm_{\mu\nu\rho}= \mathcal O(C^\pm_{\mu\nu\rho} \Lambda_\chi m_N/(4\pi)) \simeq  \mathcal O(C^\pm_{\mu\nu\rho} \Lambda_\chi  F_\pi).$ 
Because the reduced coupling of the LV gluon interaction is given by $(H_{\mu\nu\rho})^R=H_{\mu\nu\rho} \Lambda_\chi$, we obtain $\tilde H_{\mu\nu\rho} = \mathcal O(\Lambda_\chi^2 H_{\mu\nu\rho})$ for the LEC appearing in Eq.~\eqref{chirallagrH}. 

Finally, we look at the electromagnetic operators in Eq.~\eqref{effDlagr}. The reduced coupling, suppressing the Lorentz indices, is given by $(e m_N^3 \breve{D}^F)^R =e \breve{D}^F \Lambda_\chi/(4\pi)$, which should be equal to the product of the reduced couplings of the LV tensor and the electromagnetic coupling, $(D^\pm)^R e^R = e D^\pm \Lambda_\chi/(4\pi)^2$. This implies that $\breve{D}^F = \mathcal O(D^\pm/(4\pi))$.

Exactly the same rules are applied to obtain the scaling of all LECs that appear in the main text. 

\section{Reduction of effective operators using the equations of motion}
\label{app:eom}
If an effective operator vanishes when the fields satisfy the lowest-order equations of motion, i.e. when they are ``on-shell'', then a field redefinition exists that removes the operator from the Lagrangian, without changing the terms of equal or lower order \cite{PolArzt}. Therefore, to that order, the original operator does not contribute to the $S$ matrix and may be omitted from the Lagrangian entirely. We use this to show that the following operators are redundant:
\begin{subequations}
\begin{eqnarray}
\mathcal{O}_1 &=& \bar{N}T(x)\sigma^{\nu\tau}\mathcal{D}_\tau\mathcal{D}^\mu \mathcal{D}^\rho N\ , \\
\mathcal{O}_2 &=& \epsilon^{\nu\rho\alpha\beta}\bar{N}T(x)\gamma_\alpha\mathcal{D}_\beta\mathcal{D}^\mu N\ , \\
\mathcal{O}_3 &=& \epsilon^{\rho\nu\alpha\beta}\bar{N}T(x)\gamma_\alpha \gamma^5\mathcal{D}_\beta\mathcal{D}^\mu N\ , \\
\mathcal{O}_4 &=& \bar{N}T(x)\gamma_\tau\gamma^5\mathcal{D}^\tau\mathcal{D}^\mu\mathcal{D}^\nu\mathcal{D}^\rho N\ , \\
\mathcal{O}_5 &=& \bar{N}T(x)\gamma^5 \sigma^{\mu\lambda}\mathcal{D}_\lambda\mathcal{D}^\nu\mathcal{D}^\rho N\ ,
\end{eqnarray}
\label{nonphysoperators}%
\end{subequations}
where $T(x)$ represents a general operator consisting of isospin matrices, pion fields, and covariant derivatives of pion fields. With the correct choice for $T(x)$, all operators in Eqs.~\eqref{nonphysoperators} have symmetry properties corresponding to one of the LV operators in Eqs.~\eqref{lagr1gevQ} and \eqref{lagr1gevG}. If $T(x)$ does not contain derivatives of pion or photon fields, these operators naively contribute at the same order as the operators in the chiral effective Lagrangians in the main text. However, we show that they are equivalent to operators already present in these Lagrangians and/or operators that are of higher order. In Table~\ref{tab:Tforms} we summarize the forms of $T(x)$ that correspond to the dominant ($\Delta = -1$) redundant operators for terms with $C^\pm_{\mu\nu\rho}$, $D^\pm_{\mu\nu\rho}$, and $H_{\mu\nu\rho}$. 
\begin{table}[t!]
\centering
\begin{tabular}{c|ccccc}
\hline\hline
 $T(x)$ & $\mathcal{O}_1$ & $\mathcal{O}_2$ & $\mathcal{O}_3$ & $\mathcal{O}_4$ & $\mathcal{O}_5$\\
\hline
{\bf 1} & $C^+_{\mu\nu\rho}$ & $D^+_{\mu\nu\rho}$ & $C^+_{\mu\nu\rho}$ & $H_{\mu\nu\rho}$ & $H_{\mu\nu\rho}$ \\
$\tau_3 - \frac{2}{F_\pi^2 D}(\bpi^2 \tau_3 - \pi_3 \btau\cdot\bpi)$ & $C^-_{\mu\nu\rho}$ & $D^-_{\mu\nu\rho}$ & $C^-_{\mu\nu\rho}$ & - & - \\
$\frac{1}{F_\pi D}(\btau\times \bpi)_3$ & $D^-_{\mu\nu\rho}$ & $C^-_{\mu\nu\rho}$ & $D^-_{\mu\nu\rho}$ & - & - \\
\hline\hline
\end{tabular}
\caption{The expressions in Eqs.~\eqref{nonphysoperators} give different operators, depending on the form of $T(x)$. We give three possible forms of $T(x)$ and list the LV tensors to which the resulting operator corresponds. All of these are shown to be redundant.}
\label{tab:Tforms}%
\end{table}

The zeroth-order equation of motion for a nucleon field reads
\begin{eqnarray}
i\slashed{\mathcal{D}} N &=& m_N N\ .
\end{eqnarray}
Multiplying this equation by $\gamma^\mu$, we get that
\begin{equation}
m_N \gamma^\mu N = i\mathcal{D}^\mu N + \sigma^{\mu\nu}\mathcal{D}_\nu N\ ,
\label{eomrewritten}
\end{equation}
which in some cases is a more convenient form. 
After writing out the gamma matrix commutator, partial integration, and using the equation of motion and its complex conjugate, the operator $\mathcal{O}_1$ can be written as
\begin{eqnarray}
\mathcal{O}_1 &\stackrel{\rm on\ shell}{\longrightarrow}& \frac{i}{2}\bar{N}(\mathcal{D}_\tau T(x))\gamma^\tau\gamma^\nu\mathcal{D}^\mu\mathcal{D}^\rho N + \dots\ ,
\label{sigmaform}
\end{eqnarray}
where, here and in the following, the dots represent total-derivative terms and/or terms containing the electromagnetic field strength and higher-order pion terms. The latter two originate from the commutator of covariant derivatives given by
\begin{equation}
[\mathcal{D}_\mu,\mathcal{D_\nu}]N = \frac{ie}{2}F_{\mu\nu}(1+\tau_3)N + {\rm pion\ terms}\ .
\end{equation}
All terms in this expression raise the chiral index by two with respect to terms without the commutator. For convenience, we have defined in Eq.~\eqref{sigmaform}
\begin{equation}
\mathcal{D}_\mu T(x) = \partial_\mu T(x) + \frac{i}{F_\pi^2}(\bpi\times D_\mu \bpi)_a[\tau_a,T(x)] + \tfrac{ie}{2}A_\mu [\tau_3, T(x)]\ .
\end{equation}
Since $T(x)$ only contains pion or photon fields, its derivative will always either vanish or contain derivatives of these fields, which are considered to be of higher chiral order as well. Therefore, $\mathcal{D}_\mu T(x)$ raises the chiral order of the operator by one and operators of the form of $\mathcal{O}_1$ do not contribute at lowest order. The explicit expressions for $\mathcal{D}_\mu T(x)$ are given at the end of this appendix. 

Eq.~\eqref{eomrewritten} can be used to write $\mathcal{O}_2$ as
\begin{eqnarray}
\mathcal{O}_2 &\stackrel{\rm on\ shell}{\longrightarrow}& \frac{1}{m_N}\epsilon^{\nu\rho\alpha\beta}\bar{N}T(x) \mathcal{D}_\beta \mathcal{D}^\mu(i\mathcal{D}_\alpha + \sigma_{\alpha\lambda}\mathcal{D}^\lambda) N \notag \\
&=& \frac{1}{m_N}\epsilon^{\nu\rho\alpha\beta}\bar{N}(\mathcal{D}^\lambda T(x))\gamma_\lambda \gamma_\alpha \mathcal{D}_\beta \mathcal{D}^\mu N +\dots\ ,
\end{eqnarray}
where the equality is a consequence of the antisymmetry of the Levi Civita tensor. It makes the first term on the first line proportional to the electromagnetic field strength, which is of higher order. The remaining term is equal to Eq.~\eqref{sigmaform} and the equality follows.

Next, we consider operators of the form of $\mathcal{O}_3$. Such operators turn out to be equivalent to operators of the form $\bar{N}T(x)\sigma^{\nu\rho}\mathcal{D}^\mu N$, which are already present in the effective Lagrangians of Section~\ref{Npioperators}. This can be shown, using Eq.~\eqref{eomrewritten}, since
\begin{eqnarray}
\mathcal{O}_3 &\stackrel{\rm on\ shell}{\longrightarrow}& -\frac{1}{m_N}\epsilon^{\rho\nu\alpha\beta}\bar{N}T(x)\gamma^5 \mathcal{D}_\beta\mathcal{D}^\mu (i\mathcal{D}_\alpha + \sigma_{\alpha\lambda}\mathcal{D}^\lambda) N \notag \\
&=& \frac{i}{m_N}\bar{N}T(x)\left[\sigma^{\nu\tau}\mathcal{D}_\tau \mathcal{D}^\mu \mathcal{D}^\rho - \sigma^{\rho\tau}\mathcal{D}_\tau \mathcal{D}^\mu \mathcal{D}^\nu\right]N + \frac{i}{m_N}\bar{N}T(x)\sigma^{\rho\nu}\mathcal{D}^2 \mathcal{D}^\mu N + \dots \ . \notag \\
\end{eqnarray}
The first two of these operators are shown to be redundant in Eq.~\eqref{sigmaform}, while the last one is equal to the operator already present in the Lagrangian, if we use that $\mathcal{D}^2N = -m_N^2 N$ up to the current order.

For operators $\mathcal{O}_4$, using the equation of motion, we get on the one hand that
\begin{equation}
\mathcal{O}_4 \stackrel{\rm on\ shell}{\longrightarrow} im_N \bar{N}T(x)\gamma^5\mathcal{D}^\mu\mathcal{D}^\nu\mathcal{D}^\rho N + \dots\ .
\end{equation}
On the other hand, if we use partial integration to let the covariant derivative act on $\bar{N}$, we get that
\begin{equation}
\mathcal{O}_4 \stackrel{\rm on\ shell}{\longrightarrow} -im_N \bar{N}T(x)\gamma^5\mathcal{D}^\mu\mathcal{D}^\nu\mathcal{D}^\rho N -\bar{N}(\mathcal{D}_\tau T(x))\gamma^\tau\gamma^5\mathcal{D}^\mu\mathcal{D}^\nu\mathcal{D}^\rho N + \dots\ .
\end{equation}
Therefore, on-shell and up to terms of higher chiral order and a total derivative, $\mathcal{O}_4 = -\mathcal{O}_4 =0$.

For operators of the form $\mathcal{O}_5$, the equation of motion gives
\begin{eqnarray}
\mathcal{O}_5 & \stackrel{\rm on\ shell}{\longrightarrow} & \bar{N}T(x)\gamma^5 \mathcal{D}^\nu \mathcal{D}^\rho(m_N\gamma^\mu - i\mathcal{D}^\mu)N + \dots\ .
\end{eqnarray}
Using partial integration to let the covariant derivative in  $\sigma^{\mu\lambda}\mathcal{D}_\lambda$ act on $\bar{N}$, then using the equation of motion and one more partial integration, we can also write $\mathcal{O}_5$ as
\begin{eqnarray}
\mathcal{O}_5 & \stackrel{\rm on\ shell}{\longrightarrow} & \bar{N}T(x)\gamma^5\mathcal{D}^\nu\mathcal{D}^\rho(m_N\gamma^\mu + i\mathcal{D}^\mu)N - \bar{N}(\mathcal{D}^\mu T(x))\gamma^5\mathcal{D}^\nu\mathcal{D}^\rho N \notag \\
&& + \bar{N}(\mathcal{D}_\lambda T(x))\gamma^5\sigma^{\mu\lambda}\mathcal{D}^\nu \mathcal{D}^\rho N + \dots\ .
\end{eqnarray}
Combining these two results, we see that $\mathcal{O}_5$ is equivalent to
\begin{eqnarray}
\mathcal{O}_5 & \stackrel{\rm on\ shell}{\longrightarrow} & m_N\bar{N}T(x)\gamma^5 \gamma^\mu \mathcal{D}^\nu\mathcal{D}^\rho N - \bar{N}(\mathcal{D}^\mu T(x))\gamma^5\mathcal{D}^\nu\mathcal{D}^\rho N \notag \\
&& + \bar{N}(\mathcal{D}_\lambda T(x))\gamma^5\sigma^{\mu\lambda}\mathcal{D}^\nu \mathcal{D}^\rho N + \dots\ .
\end{eqnarray}
The first term is equal to an operator that is already included in the effective chiral Lagrangian, while the other terms are of higher order. In addition to $H_{\mu\nu\rho}$, as mentioned in Table~\ref{tab:Tforms}, $\mathcal{O}_5$ also corresponds to operators that could be included in Eq.~\eqref{effDlagr}. This is the case if we replace $T(x)$ by a contraction of $D^-_{\mu\nu\rho}$ and $F_{\alpha\beta}$. Using the analysis above it is then easy to show that such operators are equivalent to the one displayed in Eq.~\eqref{effDlagr}.

Finally, we give the explicit expressions for $\mathcal{D}_\mu T(x)$ for the forms of $T(x)$ in Table~\ref{tab:Tforms}:
\begin{subequations}
\begin{eqnarray}
\mathcal{D}_\mu ({\bf 1}) &=& 0\ , \\
\mathcal{D}_\mu\left(\tau_3 - \frac{2}{F_\pi^2 D}(\bpi^2 \tau_3 - \pi_3 \btau\cdot\bpi)\right) &=& \frac{4}{F_\pi^2 D}\left((D_\mu \pi)_3 \btau\cdot \bpi - \bpi\cdot D_\mu \bpi \tau_3\right)\ , \\
\mathcal{D}_\mu\left(\frac{1}{D}(\btau\times \bpi)_3\right) &=& \left(1-\frac{2\bpi^2}{F_\pi^2 D}\right)(\btau\times D_\mu \bpi)_3 + \frac{2\pi_3}{F_\pi^2 D}\bpi\cdot(\btau\times D_\mu \bpi)\ .\notag \\
\end{eqnarray}
\end{subequations}
This shows that all operators in this appendix are redundant up to higher-order operators with the correct chiral transformation properties. In other words, if we would want to extend the chiral Lagrangian to include higher-order operators, we would not need to reconsider the operators in Eqs.~\eqref{nonphysoperators}.

\end{document}